\documentclass[twocolumn]{aastex631}

\newcommand\as{\bgroup\markoverwith{\textcolor[rgb]{.5, 0, .6}{\rule[0.5ex]{8pt}{1.5pt}}}\ULon}
\newcommand{\be}{\begin{eqnarray}}
\newcommand{\ee}{\end{eqnarray}}

\usepackage{amsmath}
\usepackage{amssymb}
\usepackage{natbib}
\defcitealias{Dempsey2022}{Paper I}
\defcitealias{dittmann2023}{Paper II}

\begin{document}

\title{Runaway Eccentricity Growth: A Pathway for Binary Black Hole Mergers in AGN Disks}
\author[0000-0001-7764-3627]{Josh Calcino}\email{josh.calcino@gmail.com}
\affiliation{Theoretical Division, Los Alamos National Laboratory, Los Alamos, NM 87545, USA}
\author[0000-0001-8291-2625]{Adam M. Dempsey}
\affiliation{X-Computational Physics Division, Los Alamos National Laboratory, Los Alamos, NM 87545, USA}
\author[0000-0001-6157-6722]{Alexander J. Dittmann}
\affiliation{Theoretical Division, Los Alamos National Laboratory, Los Alamos, NM 87545, USA}
\affiliation{Department of Astronomy and Joint Space-Science Institute, University of Maryland, College Park, MD 20742-2421, USA}
\author[0000-0003-3556-6568]{Hui Li}
\affiliation{Theoretical Division, Los Alamos National Laboratory, Los Alamos, NM 87545, USA}

\begin{abstract}
Binary black holes embedded within the accretion disks that fuel active galactic nuclei (AGN) are promising progenitors for the source of gravitational wave events detected by LIGO/VIRGO. Several recent studies have shown that when these binaries form they should be highly eccentric and retrograde. However, many uncertainties remain concerning the orbital evolution of these binaries as they either inspiral towards merger or disassociate. Previous hydrodynamical simulations exploring their orbital evolution have been predominantly two-dimensional, or have been restricted to binaries on nearly circular orbits. We present the first high-resolution, three-dimensional local shearing-box simulations of both prograde and retrograde eccentric binary black holes embedded in AGN disks. We find that retrograde binaries shrink several times faster than their prograde counterparts and exhibit significant orbital eccentricity growth, the rate of which monotonically increases with binary eccentricity. Our results suggest that retrograde binaries may experience runaway orbital eccentricity growth, which may bring them close enough together at pericenter for gravitational wave emission to drive them to coalescence. Although their eccentricity is damped, prograde binaries shrink much faster than their orbital eccentricity decays, suggesting they should remain modestly eccentric as they contract towards merger. Finally, binary precession driven by the AGN disk may dominate over precession induced by the supermassive black hole depending on the binary accretion rate and its location in the AGN disk, which can subdue the evection resonance and von Ziepel-Lidov-Kozai cycles. 
\end{abstract}


\section{Introduction} \label{sec:intro}

Binary black hole (BBH) mergers inside of the accretion disks of Active Galactic Nuclei (AGN) are a promising avenue to produce gravitational wave (GW) events \citep[e.g.][]{mckernan2012, bartos2017, stone2017}. 
The possibilities of repeated mergers within the central supermassive black hole's potential well and accretion from these gas-rich environments may readily produce GW events inside the pair instability mass gap \citep[e.g.][]{gerosa2021}, and may explain the non-zero effective spin measurements in LIGO/VIRGO observations \citep[e.g.][]{mckernan2023}. The presence of gas around the merger may facilitate electromagnetic counterparts to these events \citep[e.g.][]{2019ApJ...884L..50M,tagawa2023}. 
The BBHs may be captured in the AGN disks from the nuclear star cluster \citep{bartos2017}, or form in-situ via migration  \citep[e.g.][]{bellovary2016, secunda2019,2020MNRAS.494.1203M} or gas-assisted kinetic energy dissipation \citep{tagawa2020,JLi2023, rowan2023,rowan2023b, whitehead2023}. Regarding the latter, \cite{JLi2023} recently found that BBHs commonly form with large initial eccentricities and retrograde orbits, which was largely confirmed by \cite{whitehead2023}. Once formed, BBHs can be driven to coalescence by, for example, binary-single interactions \citep{leigh2018, samsing2022} or binary-disk interactions \citep{baruteau2011}. 

Over the past several years a growing body of work has explored the interactions between BBHs and AGN disks \citep[e.g.][]{baruteau2011,Dempsey2022,YLi2021,RLi2022}. The BBHs may expand or contract depending on properties of both the binary and AGN disk.
Studies of isolated binaries typically find that binaries gain angular momentum through the accretion of gas and from torques excited in the accretion disks around both black holes \citep[e.g.][]{munoz2019, tiede2020}; depending on binary and disk parameters, the binary may shed appreciable angular momentum by delivering gravitational kicks to the material at the inner edge of the circumbinary disk (CBD) \citep[e.g.][]{DR2022,DR2023}; and resonances within the CBD may play a role in the evolution of some eccentric retrograde binaries \citep{tiede2023}. Embedded BBHs experience many of the same torques \citep[e.g.][]{YLi2021, YLi2022}. 

\cite{YLi2021} found that equal mass BBHs on prograde circular orbits will expand, but will contract if on a retrograde orbit. This contradicted earlier work \citep{baruteau2011}, which found that the BBHs on prograde orbits contract. \cite{YLi2021} showed that the sink prescription and gravitational softening used, along with the resolution, can drastically affect the torque on the binary. 

Changing the thermal properties of the gas can alter the balance of torques acting on the binary, turning expansion into contraction. Circular prograde BBHs may contract if they are able to heat their circumsingle disks (CSDs) by a factor of $\sim$3 over the background AGN disk temperature \citep{YLi2022}. However the assumption of BBHs on circular orbits has been a significant limitation of many prior studies, as simulations show that newly formed binaries should have large eccentricities \citep{JLi2022, JLi2023}. Additional studies have explored eccentric and unequal mass BBHs. \cite{RLi2022} found that eccentric equal mass BBHs merge more slowly than their circular counterparts, owing to larger positive accretion torques. However, the results of that study were sensitive to the assumed radius of the accretor, with smaller sizes leading to lower accretion torques. Unequal-mass BBHs on circular orbits appear to merge more slowly at lower mass ratios \citep[$q\equiv M_2/M_1$,][]{RLi2022, RLi2023}. The disk thermodynamics are also important, with larger adiabatic indices leading to faster contraction \citep{RLi2023}.

All of the aforementioned studies share a major limitation: they were conducted in two dimensions. 
However, if the BHs are deeply embedded in the AGN disk, accretion of gas is a three-dimensional process. 
To remedy this, our group has recently begun investigating this problem with high-resolution 3D hydrodynamical simulations. 

In \citet[hereafter Paper I]{Dempsey2022}, we showed that the torques exerted onto the binary change significantly from two to three dimensions, and that circular binaries on prograde orbits can contract depending on the binary semi-major axis $a_b$. 
Most recently, \citet[hereafter Paper II]{dittmann2023} explored BBHs on circular but inclined orbits, finding that any BBH not exactly aligned or anti-aligned with the AGN disk will be reoriented to an aligned configuration and contract. 

In this paper we expand upon these previous works and study in three dimensions how eccentric BBHs evolve on both long and short timescales.
The layout of this paper is as follows. 
We present our numerical methods, our analysis techniques, and a description of binary dynamics and time-dependent evolution in Section \ref{sec:meth}. In Section \ref{sec:res} we present the results of our 3D simulations, including a description and study of the flow and morphology of the gas around the BBH, along with the long-term time-averaged and binary-phase dependent orbital evolution. We discuss these results in the context of other results in the literature, and their impact on GW events, in Section \ref{sec:disc}. We then summarize our results in Section \ref{sec:sum}

\section{Methods}\label{sec:meth}
\subsection{Numerics}
Our numerical method is identical to that presented in \citetalias{dittmann2023}, and we refer the reader to that paper for specific details. We provide a brief summary here.
We used the publicly available code \texttt{Athena++} \citep{stone2020} with the shearing box approximation \citep{hawley1995, stone1996} and coupled it with the N-body code \texttt{REBOUND} \citep{rein2012}. With this we accurately evolve arbitrary configurations of the three-body system without relying on ad hoc precession terms \citep[e.g.][]{RLi2022} or simplifying assumptions 
(e.g., \citetalias{Dempsey2022}; \citealt{whitehead2023}).

\begin{figure}
    \centering
    \includegraphics[width=\linewidth]{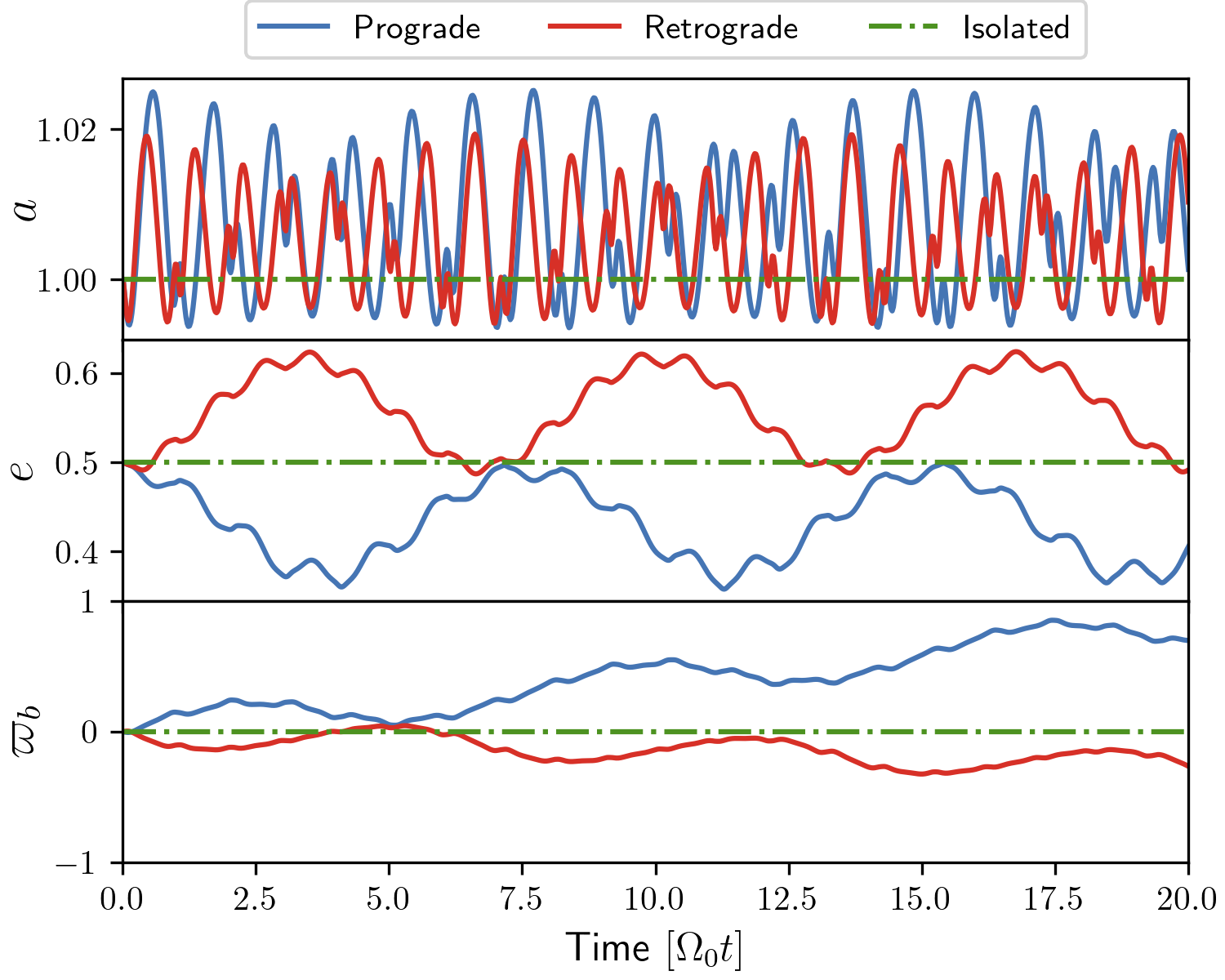}
    \caption{Evolution of the binary orbital elements for three \texttt{REBOUND} simulations which start with the same initial $a_b$, $e_b$, and $\varpi_b$. The `Prograde' and `Retrograde' curves show the binary as it orbits a central SMBH, while the `Isolated' curve is an isolated binary. Interactions with the central SMBH cause significant changes in the BBH semi-major axis and eccentricity, and induces orbital precession which is not seen in the isolated binary simulation.
    }
    \label{fig:orb_element}
\end{figure}

We study both prograde and retrograde eccentric binaries in the present work. 
In this context, prograde and retrograde refer to the relative orientation of the BBH's angular momentum vector with respect to the angular momentum vector of the binary formed between the AGN and the BBH's center of mass.
Prograde BBHs have these vectors perfectly aligned, while the vectors for retrograde BBHs are perfectly anti-aligned (for the situation without perfect alignment/anti-alignment see \citetalias{dittmann2023}). 
Figure \ref{fig:orb_element} shows the orbital elements of three binaries: a prograde binary, a retrograde binary, and an isolated binary. 
The average tidally induced precession rates for the prograde and retrograde $e_b=0.5$ binaries are approximately $\dot{\varpi}_\bullet \approx 0.1 \Omega_0$ and $-0.03 \Omega_0$, respectively.
Notably, the precession rate of the prograde and retrograde binaries oscillates over the course of each binary's orbit around the supermassive black hole (SMBH), and on average disagrees with the orbit-averaged rates applicable to more tightly bound binaries \citep[e.g.][]{2015MNRAS.447..747L,RLi2022}. Tidally effects also lead to oscillations in the binary eccentricity. The initial binary eccentricity is close to the maximum of the oscillations for the prograde binaries, but close to the minimum for the retrograde binaries. For both our prograde and retrograde initial $e_b=0$ simulations, the average eccentricity $\left< e_b \right> \approx 0.016$.

The center of our shearing-box is situated in the AGN disk at an orbital radius $R_0$, such that it orbits the central SMBH with rotation rate
$\Omega_0 = \sqrt{GM_\bullet/R_0^3}$, where $M_\bullet$ is the mass of the central SMBH. 
We work in units with $GM_\bullet=1$ and $R_0=1$, leading to $\Omega_0 = 1$. 
Our shearing box is set such that it has a constant scale height $H_0 = 0.01$. We use an isothermal equation of state with $P=c_s^2 \rho$, where the sound speed is $c_s = H_0 \Omega_0$. 
We study equal-mass BBHs with $m_b = 1.5 \times 10^{-6} M_\bullet$ such that the binary Hill radius $R_H = 0.8 H_0$ (following \citetalias{Dempsey2022} and \citetalias{dittmann2023}).
The binaries we simulate have an initial semi-major axis of $a_b = 0.25\,R_H = 0.002\,R_0$, where $R_H=(m_b/(3 M_\bullet))^{1/3} R_0$ is the radius of the Hill sphere of the binary.
We model each BH as a torque-free sink \citep{dempsey2020, dittmann2021}, with a spline-softened gravitational potential \citep{2001NewA....6...79S}; we set the sink and softening radii equal to  0.08 $a_b$.
We use initial eccentricities $e_b \in \{0.0, 0.1, 0.3, 0.5, 0.7\}$, and inclinations $i_b \in \{0, 180\}$. The binary is slowly added to the simulation over a time of $0.5\ \Omega_0^{-1}$ starting at $t = 0.5\ \Omega_0^{-1}$. 

To investigate whether long-timescale precession effects matter for embedded eccentric binaries we run each of our simulations for at least $t > 100 \Omega_0^{-1}$ -- approximately five times longer than our previous simulations. But as we show below, the simulations are converged in time after roughly $t \sim 20 \Omega_0^{-1}$ -- consistent with \citetalias{Dempsey2022} and \citetalias{dittmann2023}. 
Our simulation domain covers the region $-24H$ to $24H$ in the $x-y$ plane, and $-4H$ to $4H$ in the $z$ direction. Our base resolution is $384\times384\times64$. We utilized six levels of static mesh refinement, with our highest resolution region covering the range $-2.58\ a_b$ to $2.58\ a_b$ in the $x-y$ plane and $\pm a_b$ in the $z$ direction with a resolution of $\sim 101$ cells per $a_b$.

\subsection{Binary Evolution}

\begin{figure*}
    \includegraphics[width=1.0\linewidth]{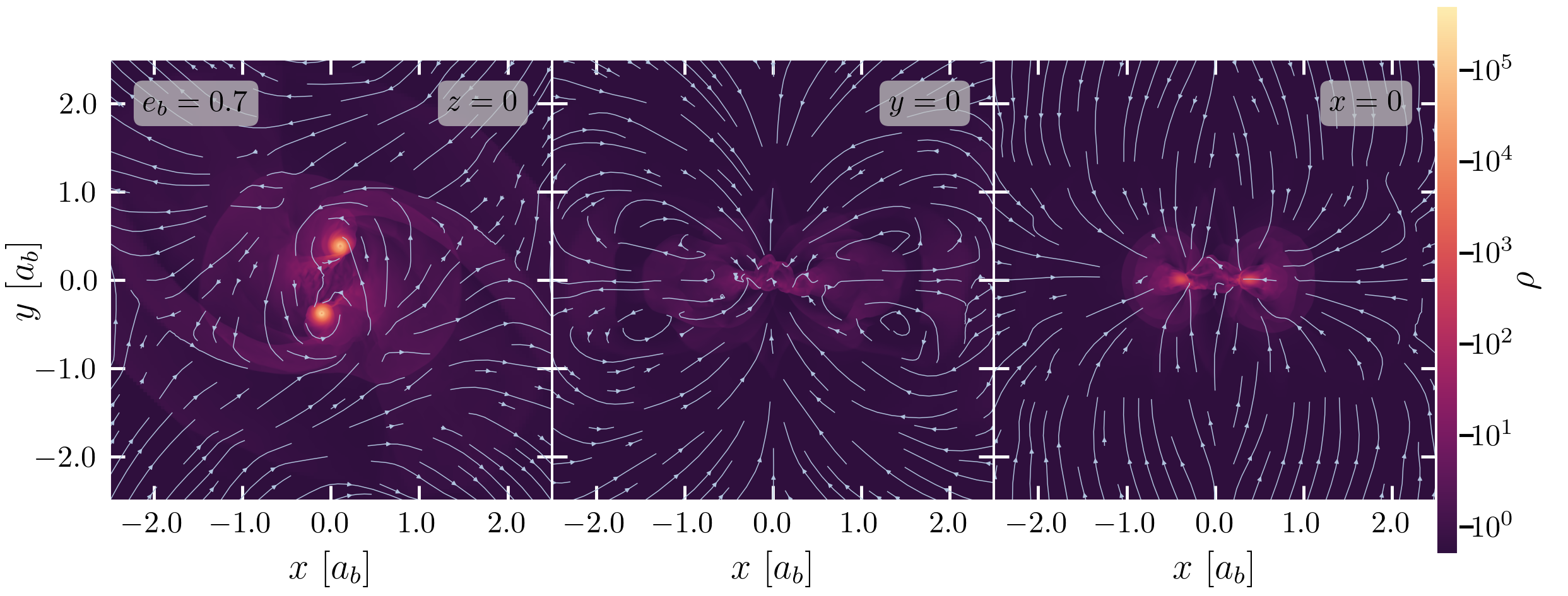}
    \caption{Density and velocity streamlines along each coordinate plane for our prograde $e_b = 0.7$ simulation. The columns show the $z=0$, $y=0$, and $x=0$ planes, in order from left to right. The light blue arrows show the velocity streamlines in their respective slice. Far from the binary the density and velocity structures are similar to an isolated BH.}
    \label{fig:xyz_slice}
\end{figure*}

\begin{figure*}
    \includegraphics[width=1.0\linewidth]{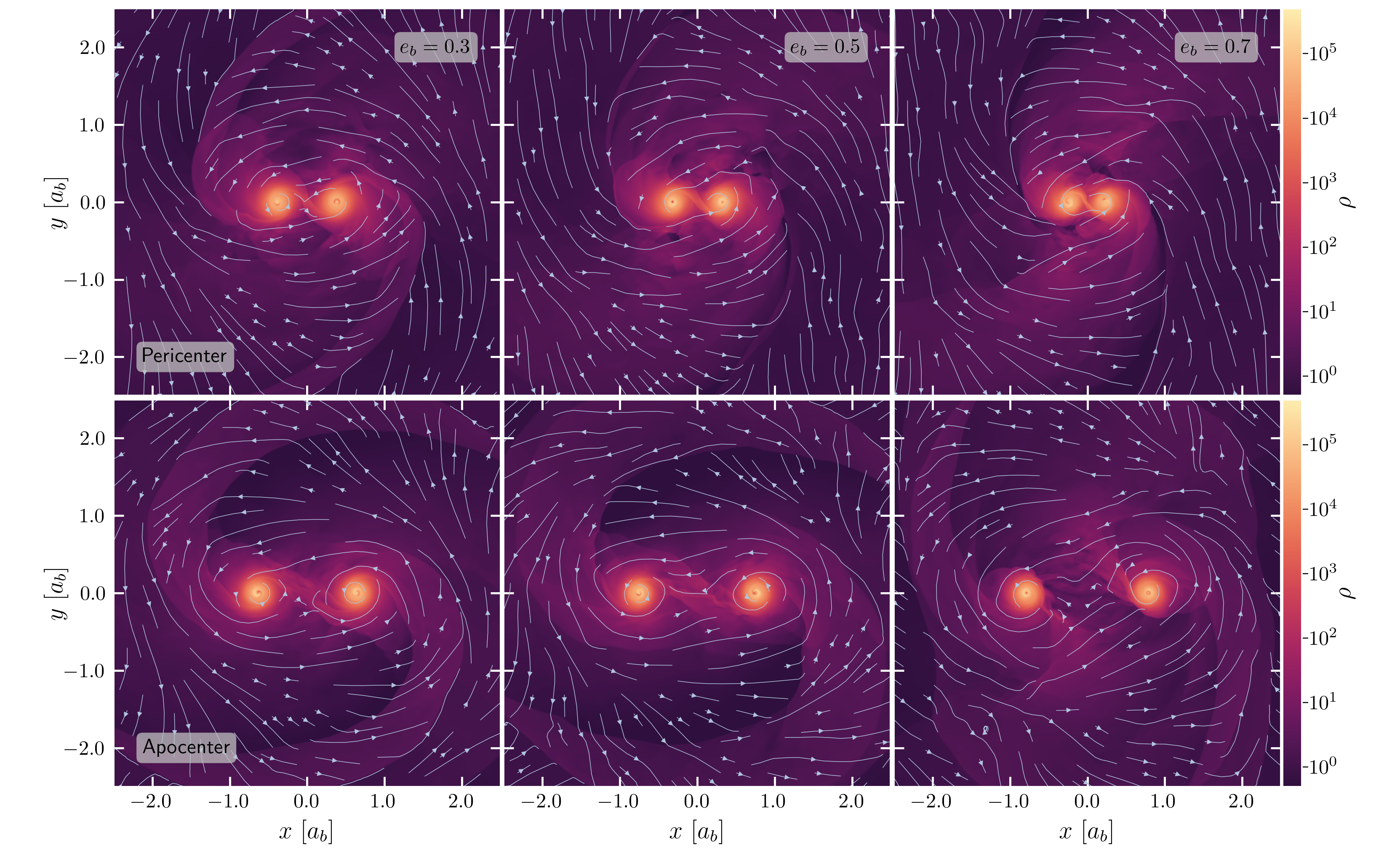}
    \caption{Density and velocity streamlines in the $z=0$ slice for our prograde simulations $e_b \in \{0.3, 0.5, 0.7\}$ for the left, middle, and right columns, respectively. The top row show the $z=0$ slice when the binary is at pericenter, while the bottom row shows the slice when the binary is at apocenter. The density and velocity structures are highly perturbed close to the binary. The inter-spiral arm between the binary increasingly becomes a transient feature at higher eccentricities.}
    \label{fig:dens}
\end{figure*}

\begin{figure*}
    \includegraphics[width=1.0\linewidth]{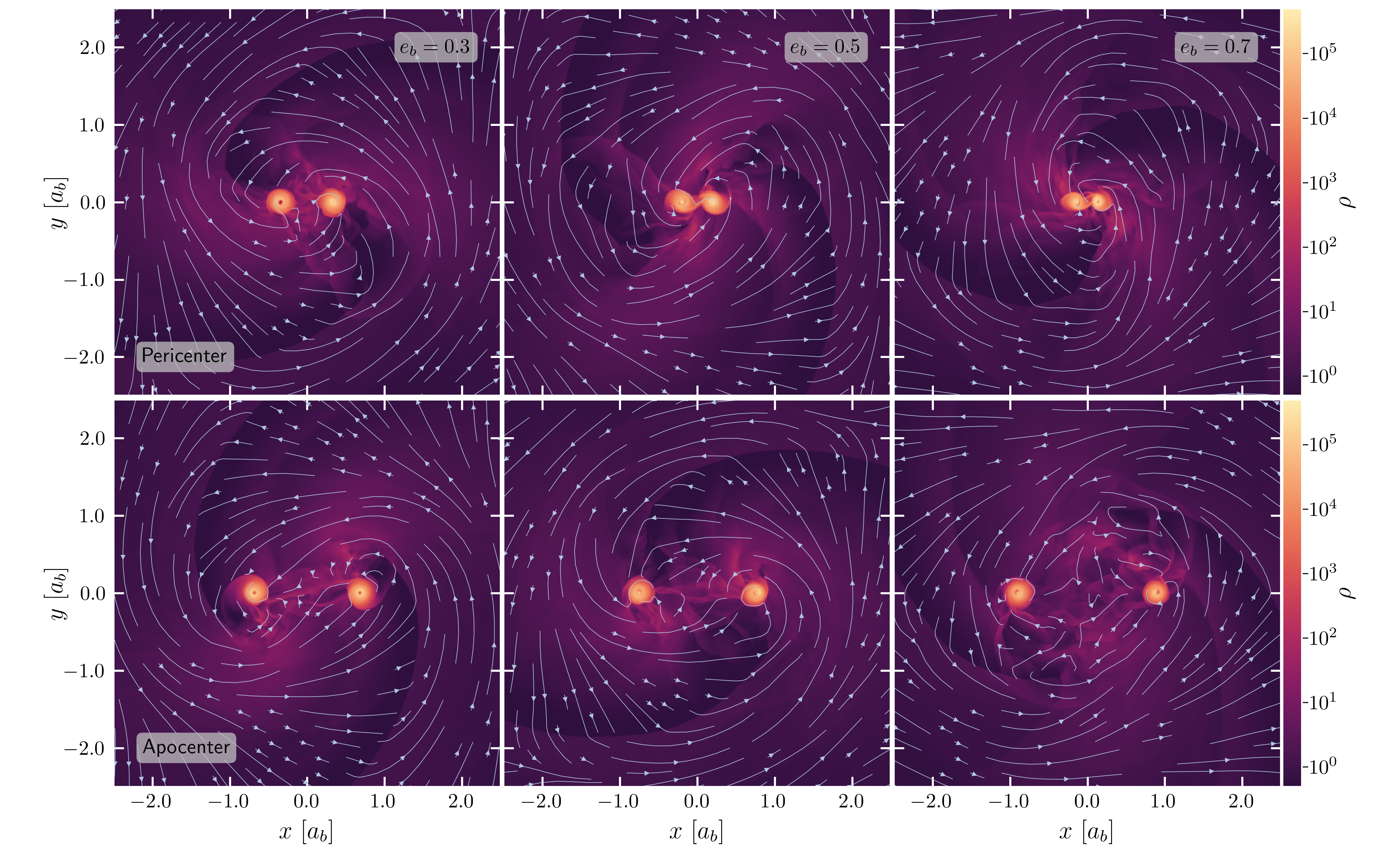}
    \caption{Same as Figure \ref{fig:dens}, but for our retrograde simulations $e_b \in \{0.3, 0.5, 0.7\}$. The CSDs around the individual BHs are much smaller than their prograde counterparts. }
    \label{fig:dens_r}
\end{figure*}

For studying the binary dynamics and evolution, we follow the prescription described in Section 2.2.1 of \citetalias{dittmann2023}, which we briefly describe here.
The specific energy, $\mathcal{E}_b$, and specific angular momentum vector, $h$, are given by 
\begin{equation}
    \mathcal{E}_b = \frac{1}{2}\bold{v}\cdot\bold{v} - \frac{Gm_b}{|\bold{r}|},
\end{equation}
and 
\begin{equation}
    \bold{h} = \bold{r}\times\bold{v},   
\end{equation}
respectively, where $\bold{v} = \bold{v_2} - \bold{v_1}$ is the relative velocity, $\bold{r} = \bold{r_2} - \bold{r_1}$, is the relative position, and $m_b = m_1 + m_2$ is the binary mass. The binary semi-major axis and eccentricity are obtained from these quantities using 
\begin{equation}
    a_b = -\frac{G m_b}{2\mathcal{E}_b},
\end{equation}
and 
\begin{equation}
    e_b^2 = 1- \frac{h^2}{G m_b a_b}.
\end{equation}
Since the central SMBH will induce procession in the BBH through tidal effects, we are also interested in the eccentricity vector 
\begin{equation}
    \bold{e} = \frac{\bold{v}\times\bold{h}}{Gm_b} - \frac{\bold{r}}{|r|}.
\end{equation}
For our co-planar binaries the longitude of periapsis $\varpi_b$ can be defined from the $x$ and $y$ components of the of the eccentricity vector, $e_x = e_b \cos\varpi_b$ and $e_y = e_b \sin \varpi_b$.
Through gravitational interactions with and accretion of the gas, the binary total energy $E = \mu_b \mathcal{E}_b$ and angular momentum $\bold{J} = \mu_b \bold{h}$, will change ($\mu_b = m_1 m_2/(m_1 + m_2)$). 
By differentiating with respect to time the above expressions for $a_b$, $e_b$, and $\varpi_b$, the following expressions can be obtained
\begin{equation}
     \frac{\dot{a_b} }{ a_b }   = \frac{\dot{m_b} }{m_b} -  \frac{\dot{\mathcal{E}_b}}{\mathcal{E}_b} ,
\end{equation} 
\begin{equation}
     \frac{\dot{e_b}^2}{1-e_b^2}   = \frac{\dot{m_b}}{m_b} -2 \frac{\dot{h}}{h} - \frac{\dot{\mathcal{E}_b}}{\mathcal{E}_b},
\end{equation}
and 
\begin{equation}
    \dot{\varpi_b} = \frac{\dot{e_y}e_x - \dot{e_x}e_y}{e_b^2}.
\end{equation}
We follow the time-averaging procedure described in Appendix A of \citetalias{dittmann2023} and normalize the quantities by the accretion timescale $m_b/\left< \dot{m_b} \right>$.

\section{Results}\label{sec:res}

\subsection{Gas Morphology}

We first examine the density and velocity structures on large scales ($\gg a_b$). In Figure \ref{fig:xyz_slice} we show the gas density and velocity streamlines for our prograde $e = 0.7$ simulation. The figure shows a mid-plane slice ($z=0$) and the two vertical slices $y=0$ and $x=0$.
As shown in \citetalias{Dempsey2022} and \citetalias{dittmann2023}, the density and velocity structures far from circular and inclined binaries are essentially identical to those of a single BH. This is also the case with our high eccentricity binaries, and is a natural consequence of the binary behaving effectively as a point mass at large distances ($r\gg a_b$). 

We show mid-plane slices of density and velocity streamlines at pericenter and apocenter for binary eccentricities $e_b \in \{ 0.3,\ 0.5,\ 0.7 \}$ in Figure \ref{fig:dens} for prograde binaries and in Figure \ref{fig:dens_r} for retrograde binaries. The density and velocity streamlines are highly perturbed close to the binary, particularly when the binary is orbiting in a retrograde sense with respect to its orbit around the central SMBH. 

We first focus on our prograde simulations. Higher binary eccentricities result in less coherent spiral structures between pericenter and apocenter. In the low eccentricity case, an inter-spiral arm connects both CSDs and its morphology does not change significantly during the binary orbit. However with increasing binary eccentricity, the inter-spiral arm becomes a transient feature, only appearing near pericenter in our $e_b = 0.7$ run, and becoming increasingly more coherent with lowering binary eccentricity. 

\begin{figure}
    \centering
    \includegraphics[width=\linewidth]{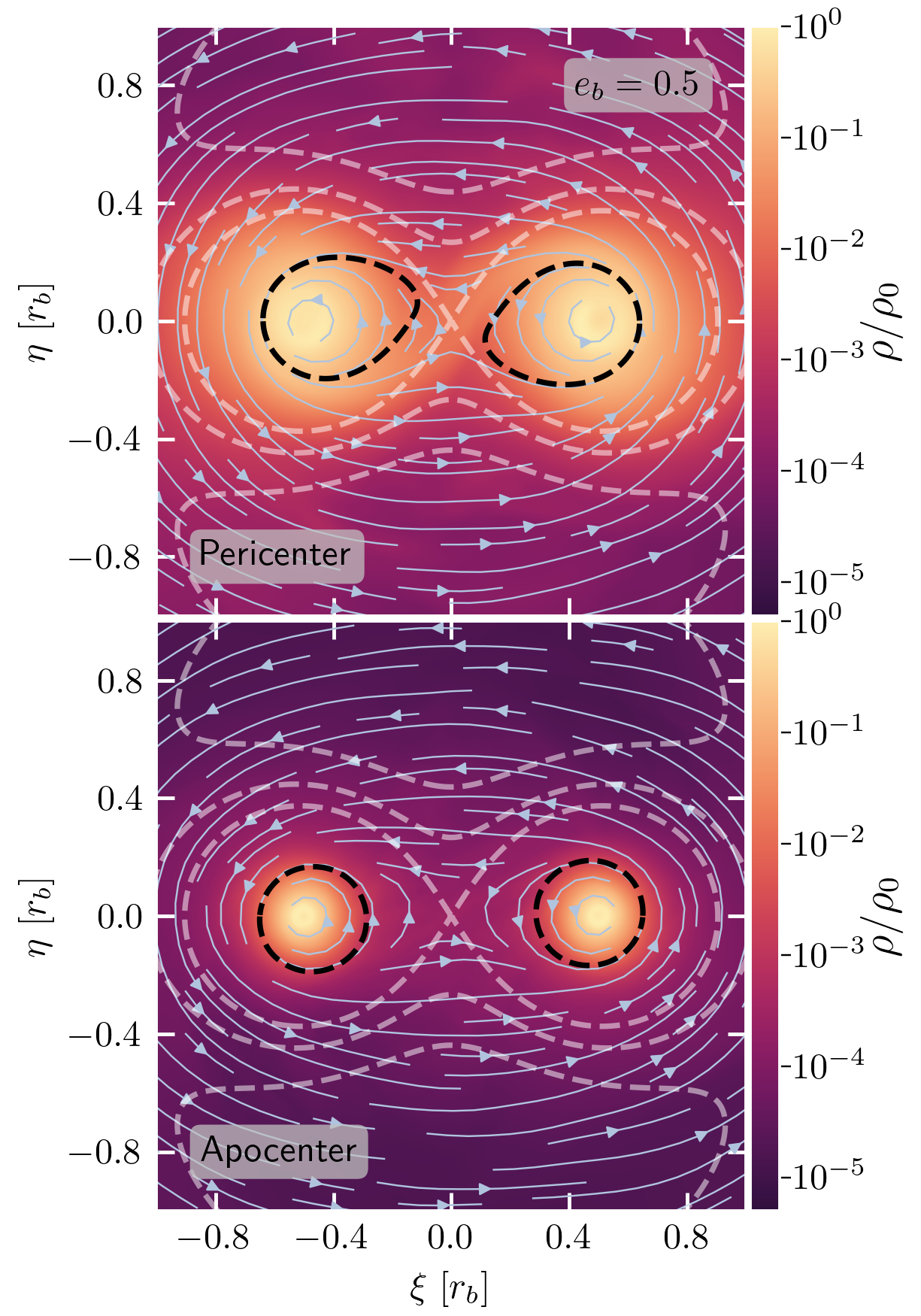}
    \caption{
    Time-averaged mid-plane slices of density and velocity at pericenter (top panel) and apocenter (bottom panel) for our prograde $e_b = 0.5$ simulation. The black dashed line shows the separatrix, and the translucent dashed-line contours show the binary potential assuming it is an isolated binary.}
    \label{fig:pro_sep}
\end{figure}

The CSDs around each individual BH change in morphology with a change in binary eccentricity and along their elliptical orbits. Higher binary eccentricities tend to produce smaller CSDs at pericenter, and the CSDs appear to increase in size as the binary orbits through its apocenter. We obtain an estimate of the size of the mid-plane CSDs by finding the velocity streamline separatrix: specifically, we determine the largest closed curve around each black hole within which all fluid should stay bound to that black hole. To remove orbit-to-orbit variations, we reran our simulations between the times of $20\ \Omega_0^{-1}$ and $26.28\ \Omega_0^{-1}$ and generate full snapshots every $0.001\ \Omega_0^{-1}$. We then averaged together each snapshot within $\pm \frac{\pi}{8}$ of the pericenter and apocenter, to obtain time-averaged midplane density and velocity profiles.

Given that the black hole positions are constantly changing, we transform our original coordinate system with a rotating-pulsating one so that the sinks lie at a common position before averaging. We use scaled, rotated coordinates \citep{munoz2019}
\begin{equation}\label{eq:xi}
    \xi = \frac{x}{r_b} \cos f_b + \frac{y}{r_b} \sin f_b
\end{equation}
and
\begin{equation}\label{eq:eta}
    \eta = \frac{y}{r_b} \cos f_b - \frac{x}{r_b} \sin f_b,
\end{equation}
where $f_b$ is the true anomaly of the binary and $r_b$ is its radial separation.
After applying this coordinate transformation, we time-averaged the slices. Figure \ref{fig:pro_sep} shows the time-averaged mid-plane density and velocity streamlines for the prograde $e_b=0.5$ simulation. 

\begin{figure}
    \centering
    \includegraphics[width=\linewidth]{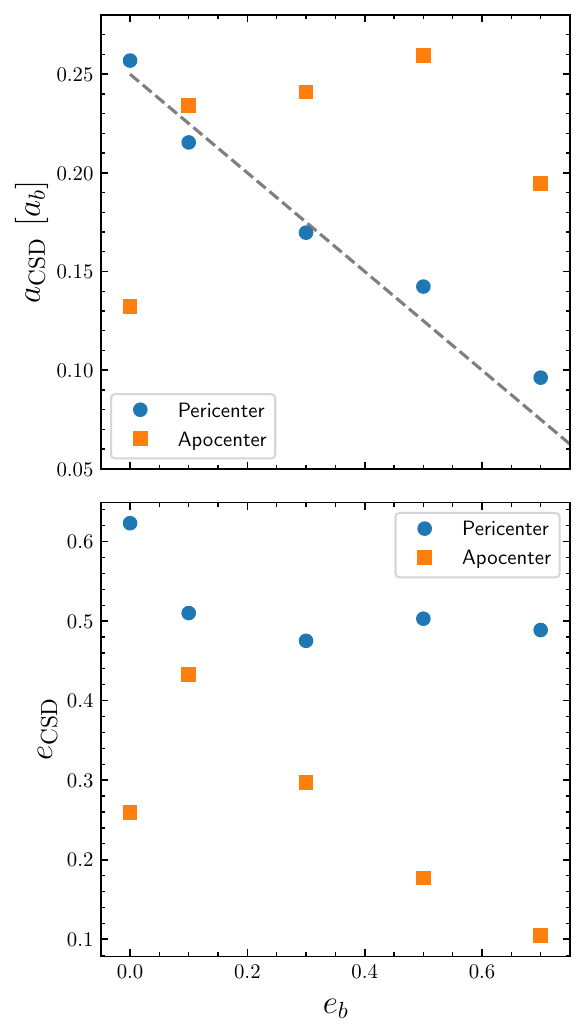}
    \caption{The fitted semi-major axis, $a_\textrm{CSD}$ (top), and eccentricity, $e_\textrm{CSD}$ (bottom), of the CSDs obtained from time-averaging mid-plane snapshots of our prograde binaries around pericenter and apocenter. The black dashed line shows the distance of the CSD to the first Lagrange point when the binary is at pericenter, which is where Roche lobe overflow will occur and is equal to half the binary separation for our equal mass binaries.}
    \label{fig:csd_fit}
\end{figure}

We find the separatrix in the time-averaged snapshots of our simulations using the method described in the Appendix. An example of the separatrix on a simulation with snapshots at peri- and apocenter is shown in Figure \ref{fig:pro_sep}, where we also plot the gas velocity streamlines and the binary potential assuming no third body interactions (i.e. we ignore the central SMBH). We applied this method of finding the separatrix to all of our prograde simulations. After finding the separatrix we fitted the path using an ellipse to obtain an estimate of the CSD semi-major axis, $a_{\rm {\tiny CSD}}$, and eccentricity, $e_{\rm {\tiny CSD}}$. These quantities are plotted as a function of binary eccentricity in Figure \ref{fig:csd_fit}, where the CSD semi-major axis is rescaled so that it is in terms of the initial binary semi-major axis. The black dashed line shows the distance of the L1 point from each BH (which is equal to half their separation) at pericenter. 

We find that CSD sizes are generally larger at aopcenter than at pericenter. At pericenter the CSD size is roughly determined by the instantaneous distance between each black hole and the binary barycenter. This is natural, as gas which departs further from either black hole may become unbound, or bound to the opposite black hole. Overall, estimating CSD size using the separatrices of velocity streamlines produces a slightly more conservative estimate than the instantaneous Roche lobe size alone. Although we measure larger CSDs at apocenter than at pericenter for binaries with $e_b>0$, they make up a smaller fraction of the binary separation. This may be due to the larger-scale flows through Hill sphere and comparatively lesser effect of the other binary component. 

\begin{figure}
    \centering
    \includegraphics[width=\linewidth]{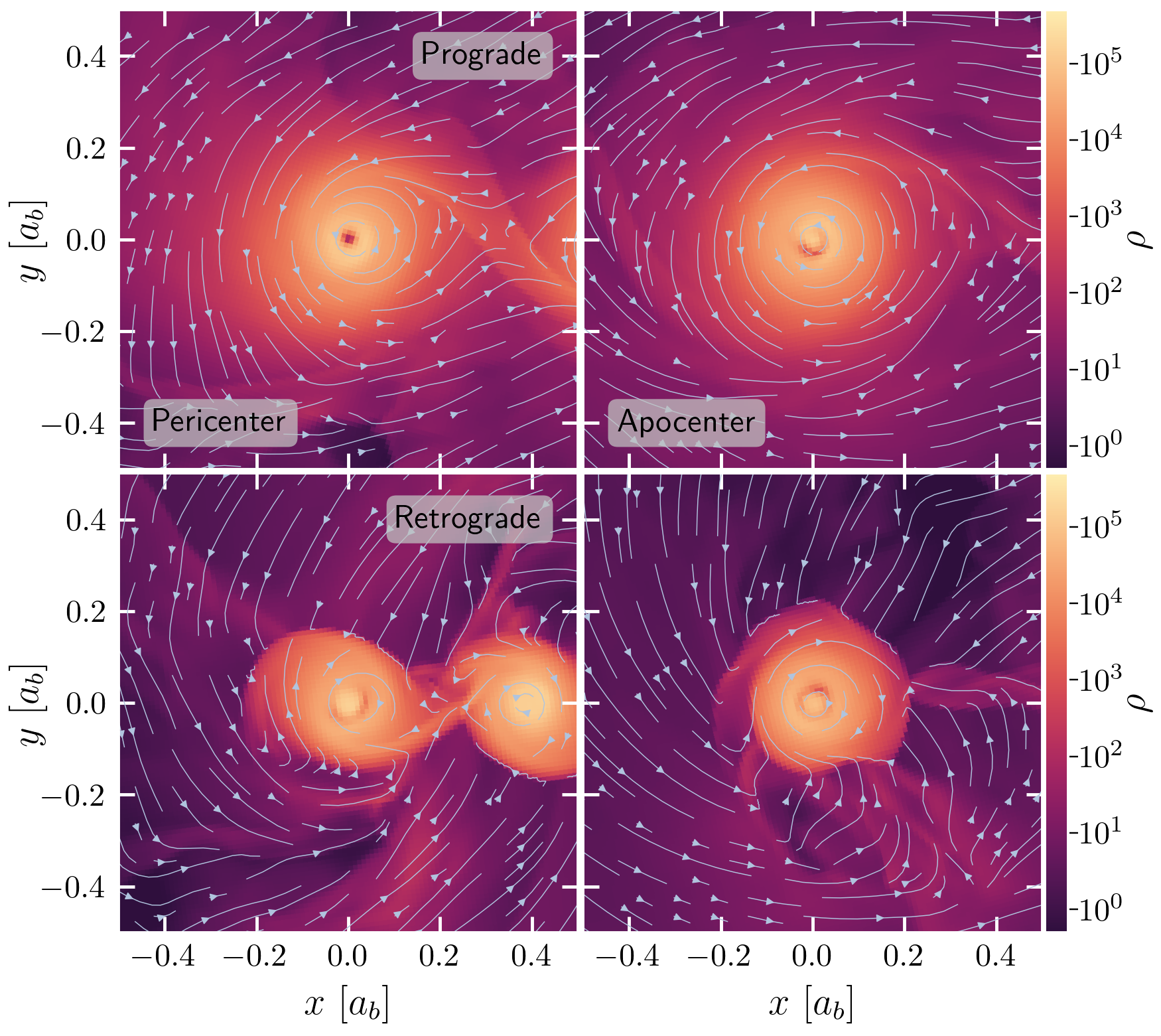}
    \caption{Mid-plane slices of our prograde (top row) and retrograde (bottom row) $e_b=0.5$ simulations at pericenter (left column) and apocenter (right column). The gas around the prograde binary orbits in a prograde manner around the binary and inside the CSD. However in the retrograde binary the gas begins as prograde and then changes sign as it falls into the CSD.}
    \label{fig:csd_slices}
\end{figure}

We turn our attention to the retrograde simulations in Figure \ref{fig:dens_r}. We observe CSDs in all of our retrograde simulations, whereas CSDs were not seen in the 2D retrograde simulation presented by \citep{RLi2022}. The CSD morphology is substantially different to that of the prograde binaries. The CSDs of the retrograde binaries all orbit in a retrograde sense with respect to the central SMBH. That is, they orbit in the same direction as the binary. We show this more clearly in Figure \ref{fig:csd_slices}, where we have plotted mid-plane slices center on a single BH for the prograde (top row) and retrograde (bottom row) $e_b = 0.5$ simulations at peri- and apocenter. We can see that the gas in the CSD of the prograde binary orbits in the same direction away from the CSD as it does inside the CSD. However in the retrograde binary the sign of the gas angular momentum flips from prograde to retrograde as the gas moves onto the CSDs.

Using the method described in the Appendix, we also found and fitted the separatrix for the retrograde simulations. 
We found, as did \citetalias{dittmann2023} for circular binaries, that retrograde CSDs are consistently smaller than their prograde counterparts (about 50\% larger than the sink radius), and the difference in size between pericenter and apocenter is much smaller. The eccentricity of the fitted ellipse to the separatrix increases monotonically with eccentricity for the CSD at pericenter, and decreases monotonically at apocenter. Given the close proximity of the CSDs to the sink and the relative insensitivity to the binary eccentricity, we are not confident that their sizes are converged. 

\subsection{Binary Orbital Evolution}
\subsubsection{Long-term Evolution}
\begin{figure}
    \centering
    \includegraphics[width=1.0\linewidth]{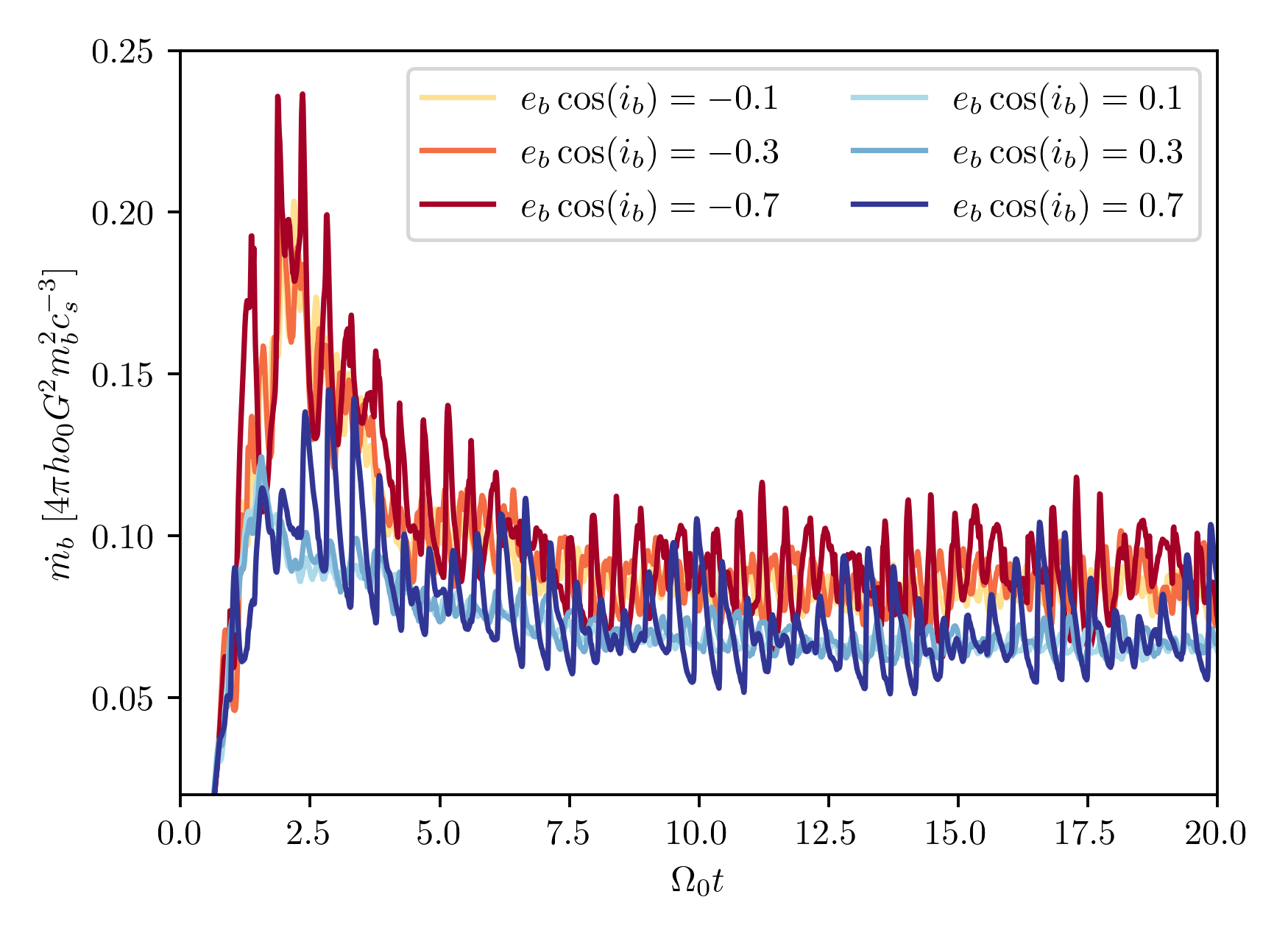}
    \caption{Binary accretion rate normalized to the Bondi rate $\dot{M_B}=4\pi \rho_0 G^2 m_b^2 c_s^{-3}$, over the first 20$\Omega_0^{-1}$. Shades of red show retrograde binaries, while shades of blue show prograde binaries. Retrograde binaries, on average, accrete more than the prograde binaries for a given initial binary eccentricity.}
    \label{fig:mdot}
\end{figure}

\begin{figure}
    \centering
    \includegraphics[width=1.0\linewidth]{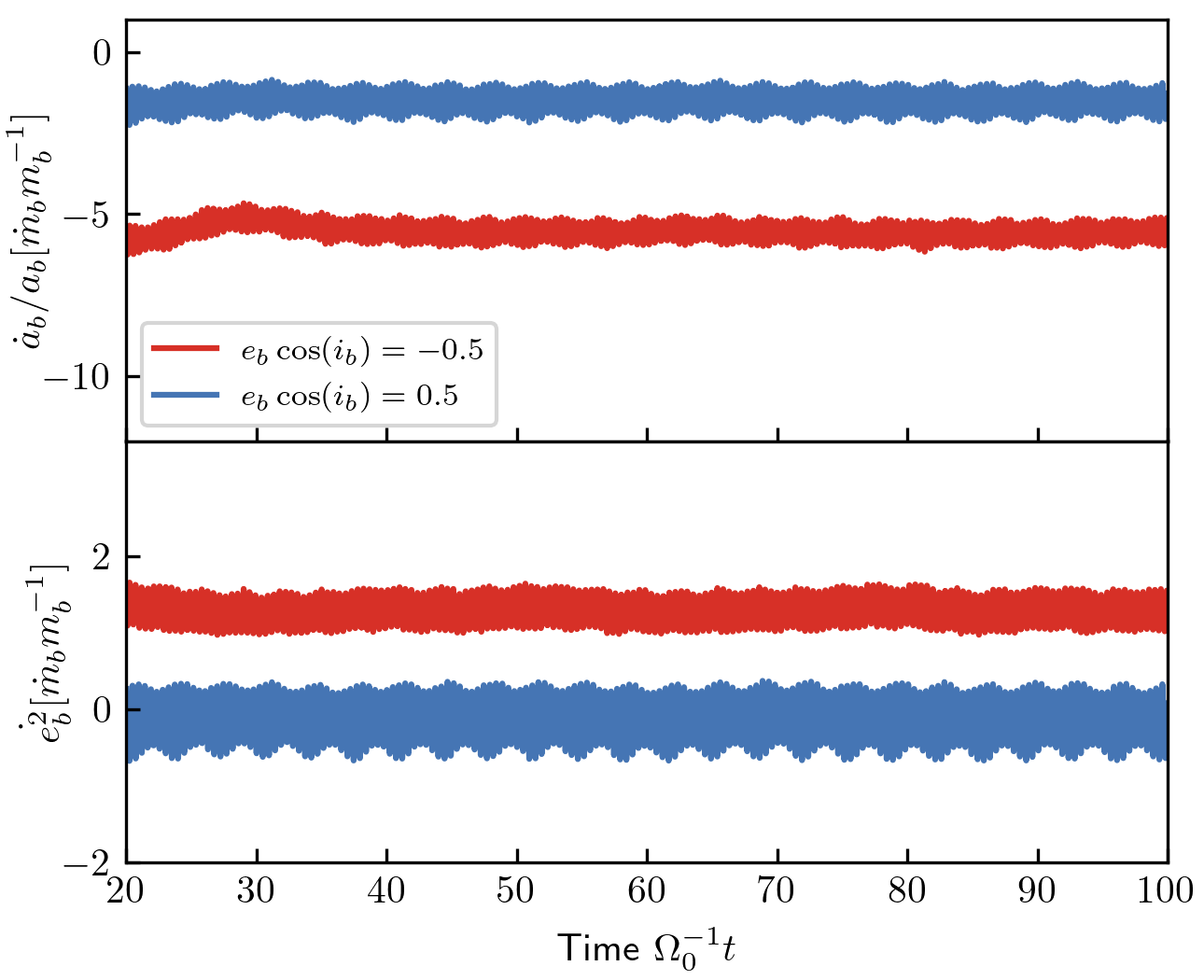}
    \caption{Time-dependent evolution of $\dot{a_b}$ and $\dot{e_b}^2$ for the prograde and retrograde $e_b = 0.5$ simulations. A moving mean with a width of $t_\textrm{width} = 2 \pi \Omega_0$ has been applied. The precession time for the prograde binary with $e_b=0.5$ is approximately $140 \Omega_0^{-1}$. No cyclic pattern in $\dot{a_b}$ and $\dot{e_b}^2$ on this timescale is seen. A small modulation with period $\Omega_0t = 2\pi$ is due to rotation of the frame of the binary as it orbits the central SMBH.}
    \label{fig:longtime}
\end{figure}

\begin{figure*}
    \centering
    \includegraphics[width=0.9\linewidth]{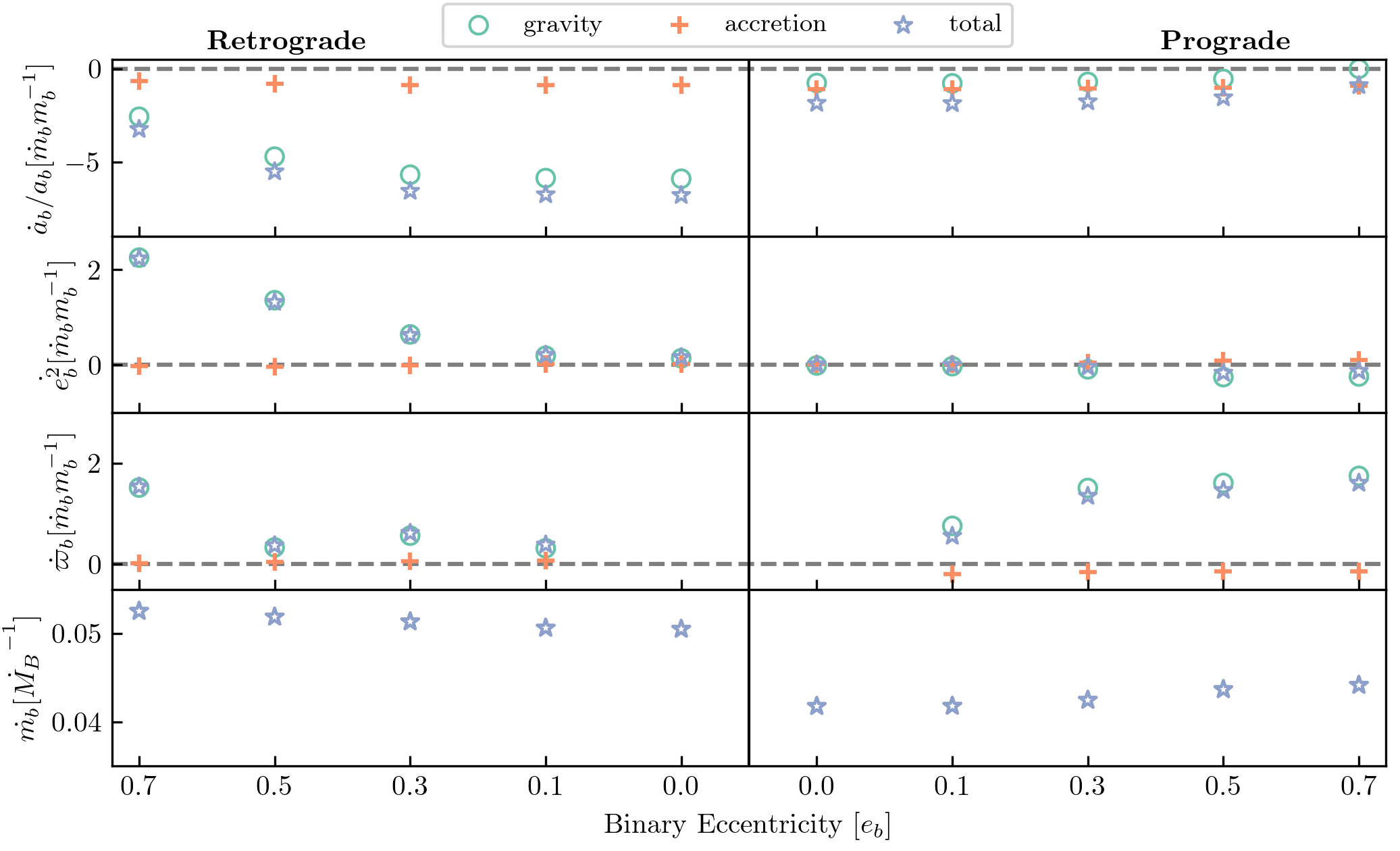}
    \caption{Time-averaged rates of change of the binary semi-major axis, eccentricity, longitude of periapsis, and the binary accretion rate for the retrograde (signified as negative eccentricity) and prograde simulations. Contributions due to gravity and accretion, as well as their sum, are shown. All quantities are time-averaged between $20\ \Omega_0^{-1}$ and $100\ \Omega_0^{-1}$. }
    \label{fig:orbital_evo}
\end{figure*}

\begin{table*}
\begin{center}
 \begin{tabular}{cccccc|ccccc}
  & \multicolumn{5}{c}{\textbf{Retrograde}} & \multicolumn{5}{c}{\textbf{Prograde}}\\
 \hline
   $e_b$ & 0.7 & 0.5 & 0.3 & 0.1 & 0.0 & 0.0 & 0.1 & 0.3 & 0.5 & 0.7 \\
 \hline
  $\dot{a_b}/a_b\ [\dot{m}_bm_b^{-1}]$ &  -3.25 & -5.53 & -6.58 & -6.76 & -6.8 & -1.84 & -1.86 & -1.76 & -1.55 & -0.889 \\
 $\dot{e_b}\ [\dot{m}_bm_b^{-1}]$ &  1.48 & 1.04 & 0.779 & 0.6 & 4.03 & 0.215 & -0.0411 & -0.115 & -0.254 & -0.279 \\
 $\dot{e_b}^2\ [\dot{m}_bm_b^{-1}]$ &  2.23 & 1.32 & 0.627 & 0.21 & 0.16 & 0.010 & -0.0023 & -0.041 & -0.167 & -0.128 \\
 $\dot{\varpi_b}\ [\dot{m}_bm_b^{-1}] $ &  1.53 & 0.368 & 0.614 & 0.383 & 409 & -657 & 0.556 & 1.35 & 1.46 & 1.61 \\
 $\dot{\varpi_{\bullet}}\,[\Omega_0]$ &  -0.0326 & -0.0302 & -0.028 & -0.0268 & -0.00111 & 0.0163 & 0.106 & 0.101 & 0.0875 & 0.0652 \\
 $\dot{m_b}\ [\dot{M_B}]$ &  0.053 & 0.052 & 0.051 & 0.051 & 0.05 & 0.042 & 0.042 & 0.042 & 0.044 & 0.044 \\
 \hline
 \end{tabular}
 \end{center}
 \caption{Time-averaged values for the evolution of the binary semi-major axis, eccentricity, gas induced precession, tidal precession, and mass accretion rate. $\dot{e_b}$ and $\dot{\varpi}_b$ are not well defined when $e_b\approx 0$.}
 \label{tab:com}
 \end{table*}

Figure \ref{fig:mdot} shows the accretion rate of the BHs as a function of time for three prograde and three retrograde simulations. We can see that after a time of $\sim 20\ \Omega_0^{-1}$ the simulations have reached quasi-steady state. Prograde binaries accrete less gas on average than retrograde binaries, as observed previously in \citetalias{dittmann2023}, but the amount they accrete increases with binary eccentricity.
Figure \ref{fig:longtime} shows the time-dependent evolution of $\dot{a_b}$ and $\dot{e_b}^2$ for the prograde and retrograde $e_b = 0.5$ simulations, where a moving mean with a width of $t_\textrm{width} = 2 \pi \Omega_0^{-1}$ has been applied. The frame of the shearing-box changes with a period of $t = 2\pi \Omega_0^{-1}$ as the binary orbits the central SMBH. This produces some of the periodic short-timescale variation in $\dot{a_b}$ and $\dot{e_b}^2$ shown in Figure \ref{fig:longtime}. 

We show the time-averaged rate of change of binary semi-major axis, eccentricity, and longitude of periapsis, along with the binary accretion rate, in Figure \ref{fig:orbital_evo} and Table \ref{tab:com}. Starting with the rate of change of the semi-major axis, we note that the prograde binaries are somewhat sensitive to the binary eccentricity, but they all contract for our choice of $a_b/R_H$ \citepalias[cf.][]{Dempsey2022}.
The retrograde binaries on the other hand contract significantly faster, by a factor of $\sim 3-4$, compared with the prograde binaries with similar eccentricities. 

The rate of change of $\varpi_b$ is generally positive for both the prograde and retrograde binaries, and increases with binary eccentricity. Gravitational interactions with the AGN disk induce much more precession than gas accretion, and cause the binary to precess in a prograde sense. Accretion onto the binary is dependent both on the binary eccentricity and the direction the binary orbit. Depending on the ambient density of the AGN disk, or equivalently in our simulations the accretion rate onto the binary, these disk-induced precession rates may exceed the precession induced by tidal interactions with the SMBH ($\dot{\varpi}_\bullet$), which are listed in Table \ref{tab:com} for comparison.

Finally, we turn to binary eccentricity evolution. In agreement with other works \citep[e.g.][]{RLi2022}, we find that the binary eccentricity is damped for prograde binaries, and the degree of damping increases with eccentricity. However this is not the case for our retrograde binaries, where we find that the binary eccentricity grows and the rate of change increases with eccentricity. This suggests that eccentric retrograde binaries may experience a runaway growth in the eccentricity.
We discuss the astrophysical implications of this result in Section \ref{sec:evol}.

\subsubsection{Phase-dependent Evolution}\label{sec:phase}

\begin{figure}
    \centering
    \includegraphics[width=1.0\linewidth]{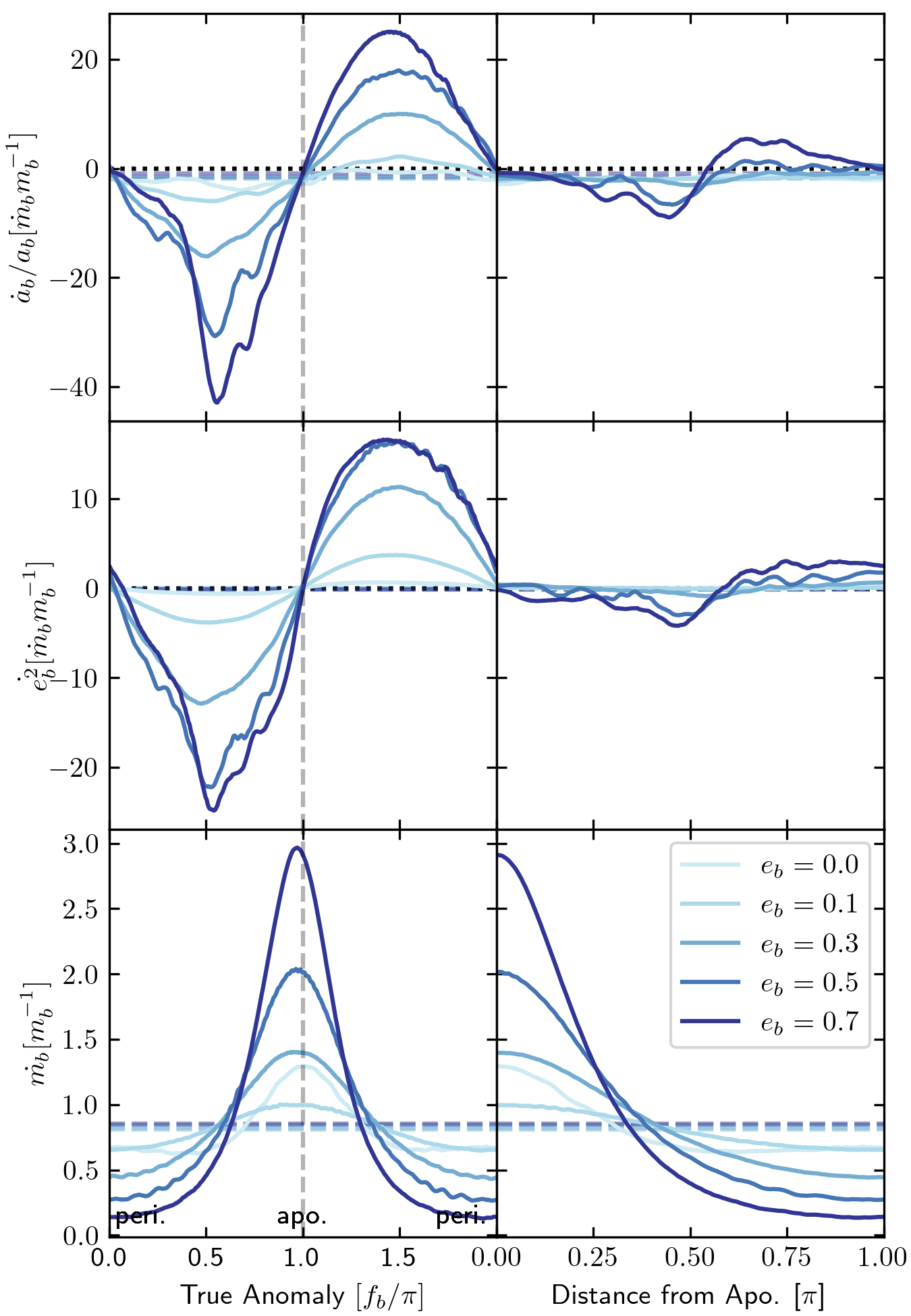}
    \caption{True anomaly folded ($f_b$-folded) binary semi-major axis rate of change $\dot{a_b}$, binary eccentricity rate of change $\dot{e_b}^2$, and binary accretion rate, for our prograde simulations.}
    \label{fig:pro_phase}
\end{figure}

\begin{figure}
    \centering
    \includegraphics[width=1.0\linewidth]{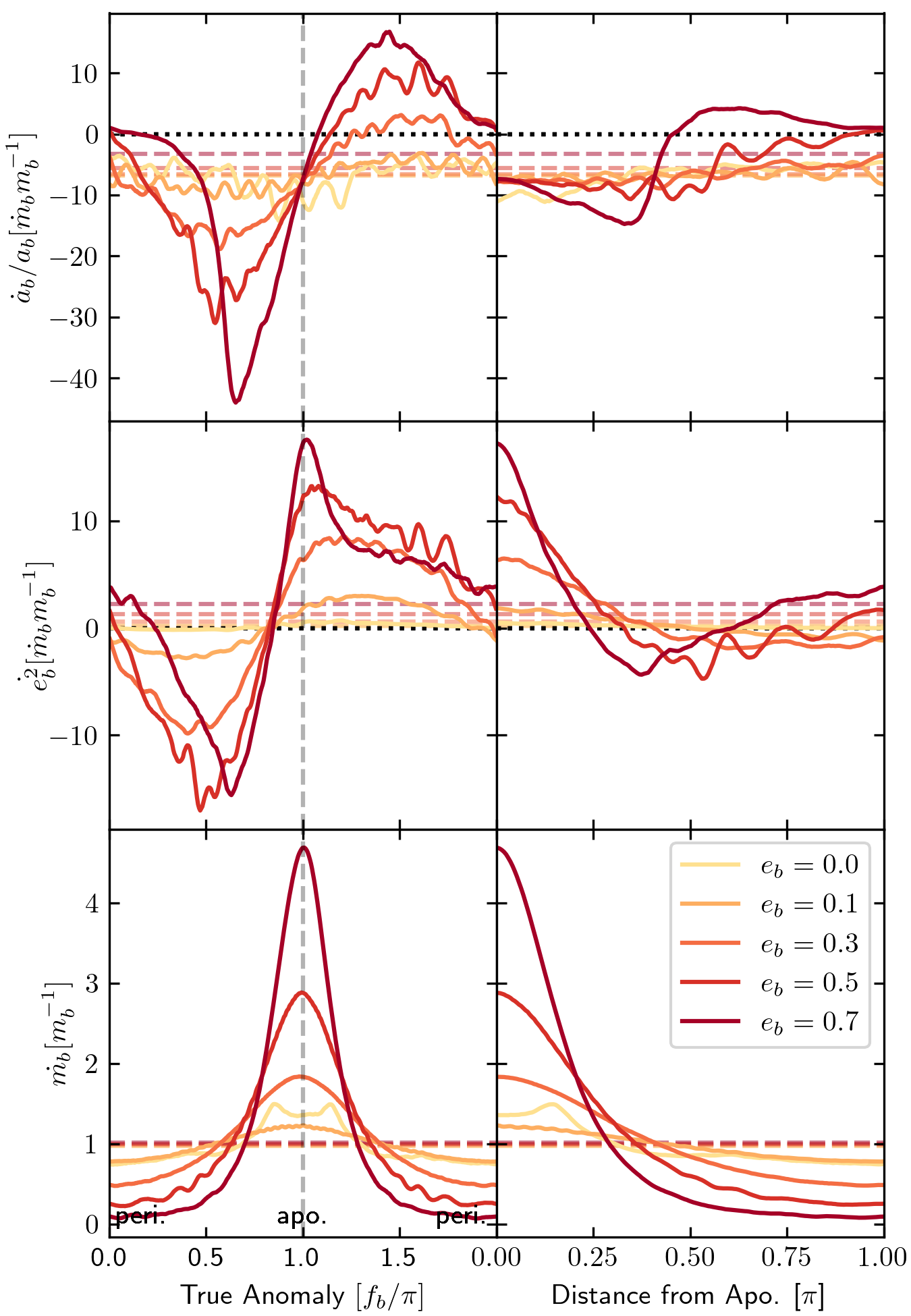}
    \caption{As in Figure \ref{fig:pro_phase}, but for the retrograde simulations.}
    \label{fig:ret_phase}
\end{figure}

The left column of Figure \ref{fig:pro_phase} shows the true anomaly ($f_b$) folded semi-major axis rate of change $\dot{a_b}$, eccentricity rate of change $\dot{e_b}^2$, and binary accretion rate $\dot{m_b}$. The solid lines show the $f_b$-dependent value of the relevant quantities, while the dashed colored horizontal line is the $f_b$-averaged quantity.\footnote{Since the phase-averaged and time-average values are exactly the same provided the average is done over the same time window, we use the terms `phase-averaged' and `time-averaged' interchangeably. The weight of each phase bin accounts for the amount of time the binary spends at each phase.} The phase-dependent curves are generated using an average shifted histogram \citep{scott1985}. 
All quantities are averaged between $20\ \Omega_0^{-1}$ and $100\ \Omega_0^{-1}$. Increasingly darker shades of blue show increasing binary eccentricity. We have added a dashed vertical line in the left column to show the apocenter, and a dotted horizontal line along $y=0$ for the $\dot{a_b}$ and $\dot{e_b}$ quantities. 
The right column shows the same quantities but we have added the region $\pi \leq f_b \leq 0$ with the region $\pi < f_b \leq 2\pi$ and divide by 2 (i.e. we take the average of the pre-apocenter evolution with the post-apocenter evolution in a way that is symmetric about the apocenter). We will denote these curves as `half-orbit-folded' for clarity. If some arbitrary quantity $g$ displays $\pi$ rotational symmetry about the point $x = 1$, $y = \langle g \rangle$ (e.g. as the function $\sin (x/\pi) + \langle g \rangle$ does) in the phase-folded (left) column of Figure \ref{fig:pro_phase}, then the corresponding curve on the half-orbit-folded (right) column will be equal to $\langle g \rangle$ everywhere.
Thus any deviations away from the phase-averaged value $\langle g \rangle$ in the half-orbit-folded plot shows that there is an asymmetry in the quantity $g$ as the binary orbits toward and away from apocenter. If there are no deviations then each part of the binary orbit is equally responsible for the non-zero phase-averaged value.

Starting with the $f_b$-folded accretion, we see that it is mostly symmetric around the apocenter, with a slight asymmetry occurring with a larger binary eccentricity. Most of the accretion occurs close to apocenter, while it drops to nearly naught at pericenter for our largest eccentricity ($e_b=0.7$). 
Interestingly, the nearly circular BBH ($e_b=0$) still shows an $f_b$-dependent accretion profile, and the peak accretion near apocenter is larger than the $e_b = 0.1$ simulation. Moving on to the binary semi-major axis evolution. As the binary moves from pericenter to apocenter, the $f_b$-dependent $\dot{a_b}$ is negative, while it is positive when the binary moves from apocenter to pericenter. We can see from the right column of Figure \ref{fig:pro_phase} that for an increasing eccentricity, the phase-dependent behavior of $\dot{a_b}$ becomes increasingly more asymmetric. The $\dot{a_b}$ slightly before and after apocenter and pericenter balance each other out. It is the region between apocenter and pericenter that produces net changes in $\dot{a_b}$ in the prograde simulations. The net effect on the binary towards apocenter is a negative $\dot{a_b}$, while closer to pericenter it is a positive $\dot{a_b}$. Overall the negative $\dot{a_b}$ wins, however with higher binary eccentricity the time-averaged $\langle \dot{a_b} \rangle$ is less negative.

The binary eccentricity evolution shows a similar trend. The $f_b$-dependent $\dot{e_b}^2$ is negative when the binary moves from pericenter to apocenter, and positive when it moves from apocenter to pericenter. Increasing binary eccentricity tends to introduce asymmetries in the phase-dependent behavior that are similar to the $\dot{a_b}$ curve. This is mostly balanced out in the phase-average, although $\langle \dot{e_b}^2\rangle$ does become more negative with larger binary eccentricity. The increasingly negative $\dot{e_b}^2$ contribution closer to apocenter outpaces the positive $\dot{e_b}^2$ component as the binary approaches pericenter.

We now turn attention to the retrograde simulations. The phase-folded binary accretion looks essentially identical to the prograde case, with the exception that the accretion rate is higher. The accretion peaks at apocenter, and drops at pericenter. The $f_b$-dependent $\dot{a_b}$ also shares similarities with the prograde case. The pericenter to apocenter phase of the binary orbit has a negative $\dot{a_b}$, while the apocenter to pericenter phase has a positive $\dot{a_b}$. However compared with the prograde simulations, there is a larger asymmetry between the pericenter to apocenter and apocenter to pericenter evolution in the $\dot{a_b}$ phase-curve. This results in a significant negative $\dot{a_b}$ at apocenter, and the positive $\dot{a_b}$ from the binary orbit leaving apocenter does not balance out the negative $\dot{a_b}$ from it approaching apocenter. This leads to a negative time-averaged $\dot{a_b}$.

The phase-folded $\dot{e_b}^2$ in the second row of Figure \ref{fig:ret_phase} is markedly different from the prograde case. There is a substantial peak in binary eccentricity growth around the apocenter, and for $e_b \in \{ 0.5, 0.7\}$, a peak at pericenter as well. The half-orbit-folded curve shows that $\dot{e_b}^2$ is slightly negative, or mostly cancels to naught, away from apocenter, but peaks again for the higher eccentricities. These negative $\dot{e_b}^2$ contributions do not completely cancel out the large positive peak of $\dot{e_b}^2$ at apocenter, resulting in the final positive $\dot{e_b}^2$.

\subsection{Torque and Power}
In the previous section we established that for all our simulations $\dot{a_b}$ is negative, while $\dot{e_b}^2$ is negative for prograde binaries, and is positive for retrograde binaries. In this section we pursue a more detailed understanding of the torque and power contributions that are responsible for these results.

\begin{figure}
    \centering
    \includegraphics[width=1.0\linewidth]{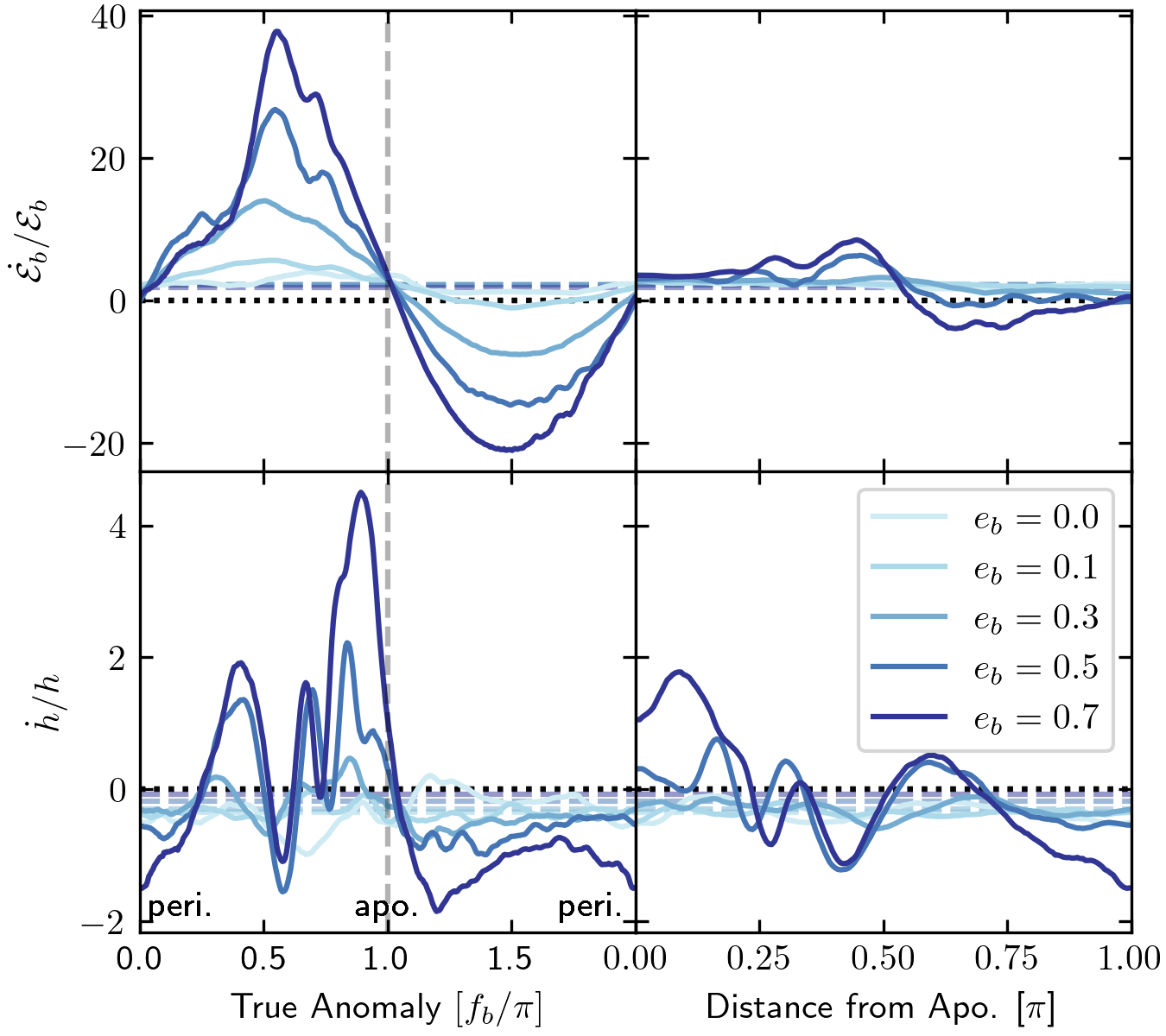}
    \caption{True anomaly folded ($f_b$-folded) normalized torque $\dot{h}/h$, and normalized power $\dot{\mathcal{E}_b}/\mathcal{E}_b$, for our prograde simulations.}
    \label{fig:pro_l0phase}
\end{figure}

\begin{figure}
    \centering
    \includegraphics[width=1.0\linewidth]{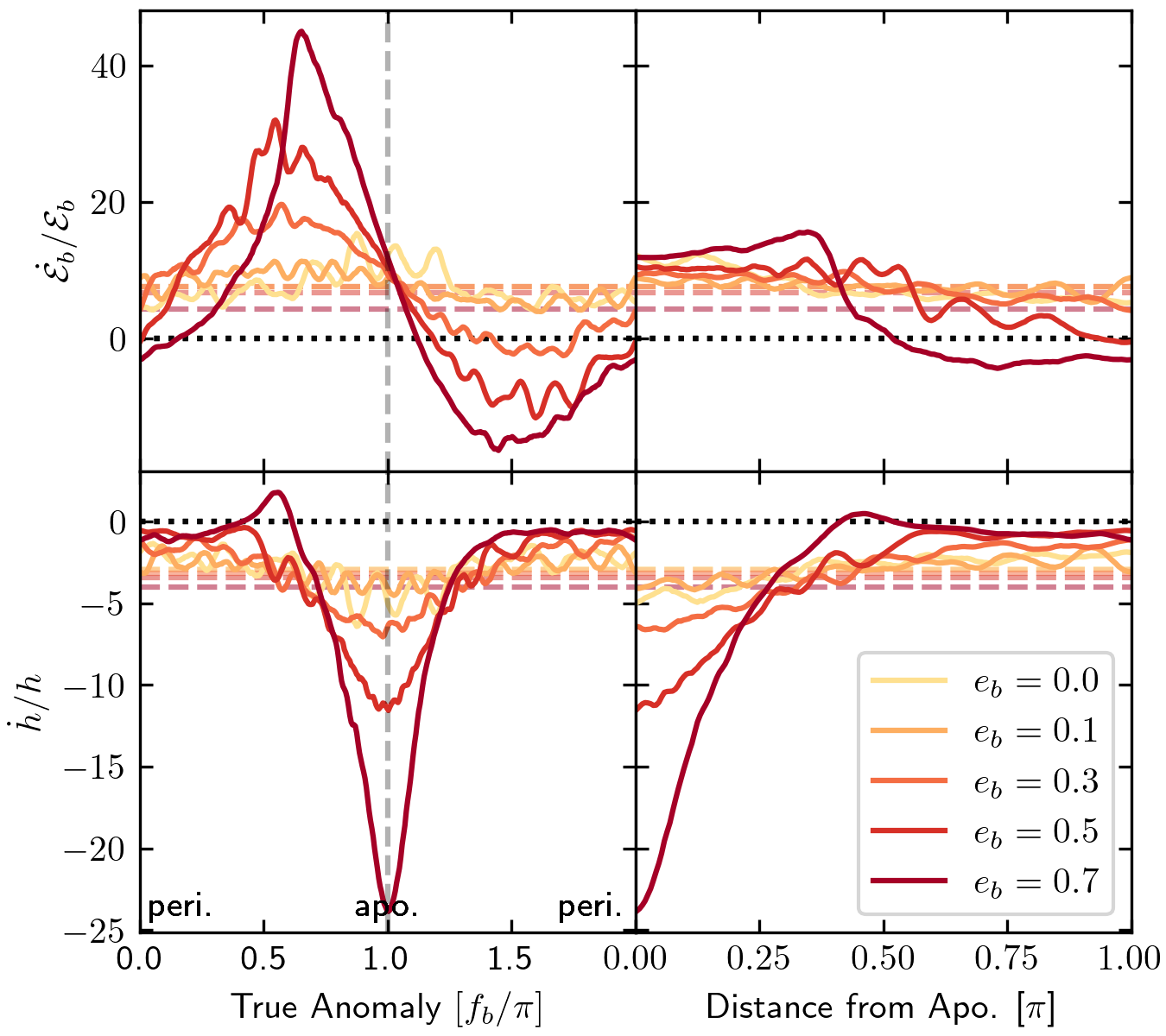}
    \caption{As in Figure \ref{fig:pro_phase}, but for the retrograde simulations.}
    \label{fig:ret_l0phase}
\end{figure}

We begin with the phase-folded of $\dot{\mathcal{E}_b}/\mathcal{E}_b$ and $\dot{h}/h$ in Figure \ref{fig:pro_l0phase} and Figure \ref{fig:ret_l0phase} for the prograde and retrograde simulations, respectively. Recalling that $\dot{a_b} = \dot{m_b}/m_b-\dot{\mathcal{E}_b}/\mathcal{E}_b$, the interpretation of how the power phase-folded profile relates to the $\dot{a_b}$ phase-folded profile is straight forward. Between the prograde and retrograde simulations, $\dot{\mathcal{E}_b}/\mathcal{E}_b$ looks quite similar. The torque, on the other hand, is drastically different between the prograde and retrograde simulations. The prograde simulations have, on average, a positive $\dot{h}/h$ from pericenter to just after apocenter, while $\dot{h}/h$ is mostly negative after apocenter. However $\dot{h}/h$ is negative almost everywhere in the retrograde simulations, dipping further at apocenter. Recalling that $\dot{e_b}^2 = (1-e_b^2)(\dot{m_b}/m_b - 2\dot{h}/h - \dot{\mathcal{E}_b}/\mathcal{E}_b)$, we can see that if both $\dot{\mathcal{E}_b}/\mathcal{E}_b$ and $\dot{h}/h$ are positive then $\dot{e_b}^2$ will be negative. This is the case for the prograde simulations. However if either $\dot{\mathcal{E}_b}/\mathcal{E}_b$ or $\dot{h}/h$ are sufficiently negative, then $\dot{e_b}^2$ can go from positive to negative. This occurs in the retrograde simulations where $\dot{e_b}^2\approx 0$ when $e_b \approx 0$, and $\dot{e_b}^2$ becomes increasingly more positive for larger $e_b$.

\begin{figure}
    \centering
    \vspace{1cm}
    \includegraphics[width=0.9\linewidth]{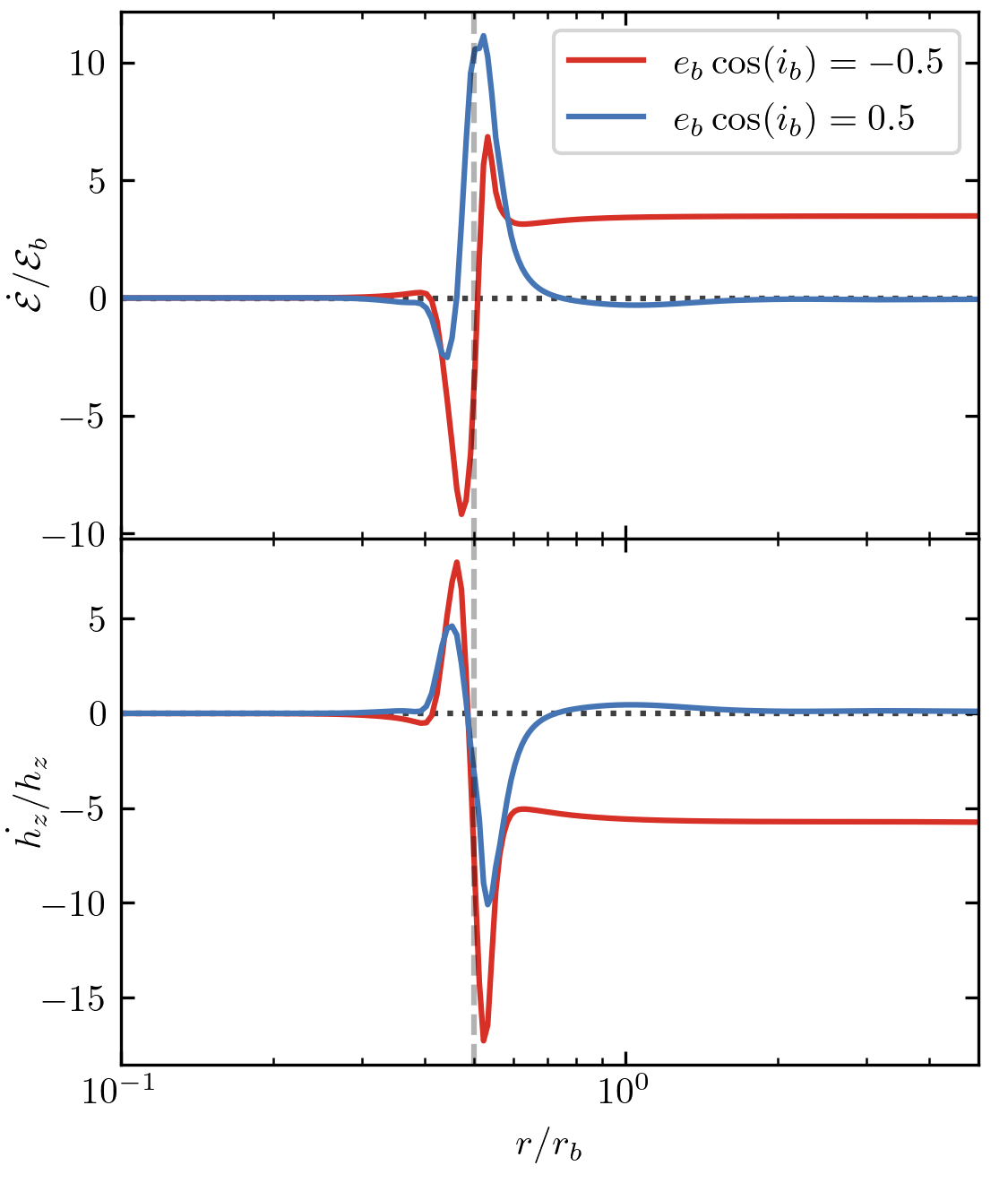}
    \caption{Time-averaged normalized power $\dot{\mathcal{E}_b}/\mathcal{E}_b$ (top) and normalized torque $\dot{h}_z/h_z$ as a function of position in $(\xi , \eta)$ space. The positions of the sink particles are marked with a gray dashed vertical line. 
    }
    \label{fig:l0_p0_apo}
\end{figure}

Overall, the binary evolution at apocenter makes the most important contribution to the retrograde binary evolution. To more adequately understand these results, we study the spatial dependence of $\dot{h}/h$ and $\dot{\mathcal{E}_b}/\mathcal{E}_b$ at apocenter. 
We compute the torque and power on the snapshots that lie in the region $\frac{7 \pi}{8} \leq f_b \leq \frac{9\pi}{8}$, then compute the cumulative 1D torque and power as a function of $r$, and average these 1D curves. The results for the prograde and retrograde $e_b = 0.5$ simulations is shown in Figure \ref{fig:l0_p0_apo}. The vertical dashed line at $r = 0.5$ shows the location of the binary.

The cumulative curve of $\dot{\mathcal{E}_b}/\mathcal{E}_b$ for the prograde simulation shows that inside the binary orbit there is a slight negative component of the power, before a sharp increase inside the CSDs. Outside of the CSDs the power drops before slightly increasing again in the CBD region. In the retrograde simulation the power increases slightly toward the CSDs but then declines inside of the CSDs. After crossing the sink region, the power then sharply increases, with a decline and then slight increase from the CSD to CBD region. Unlike the prograde case, the power does not balance out around the CSDs and the resulting power is strongly positive.

The second row of Figure \ref{fig:l0_p0_apo} shows that the torque also mostly balances out around the CSDs for the prograde $e_b = 0.5$ simulation. The CSD region produces a slight net positive torque, which is then damped slightly by the CBD region. In the retrograde simulation the torques do not balance out in the CSD region and a large negative torque is produced in the CSD region. It is clear from these profiles that the torque and power originating from the CSD region are determining the orbital evolution of the binary.

\section{Discussion}\label{sec:disc}

\subsection{Eccentricity Evolution and Orbital Damping}\label{sec:evol}

Although limited to nearly-circular binaries, \citetalias{dittmann2023} found that inclined and retrograde binaries on circular orbits have their eccentricities pumped. Eccentricity pumping of retrograde binaries has also been observed in the SPH simulations of \citep{rowan2023}, and in the study of retrograde extreme mass-ratio inspirals in \citet{2021ApJ...908L..27S}. On the contrary, no eccentricity pumping was identified in the 2D retrograde binary simulations in \citep[][]{RLi2022}. Possible reasons for this discrepancy include the two-dimensional nature of those simulations, and their prescription of a constant precession rate compared to our full solution of the three-body problem and three-dimensional hydrodynamic simulations. 

Recently, \cite{tiede2023} simulated retrograde and eccentric isolated binaries in 2D and found that retrograde binaries experience eccentricity pumping, although the eccentricity growth does not increase monotonically with binary eccentricity. That work also found that $\dot{e_b}^2 \approx 0$ when $e_b \approx 0$. We find that the eccentricity evolution $\dot{e_b}^2$ is dependent on the binary eccentricity such that higher binary eccentricity leads to a larger magnitude for $\dot{e_b}^2$. This suggests that retrograde binaries should \emph{increase} their orbital eccentricity with time, and their eccentricity growth will increase in a runaway manner until the point where gravitational waves can quench their eccentricity and shrink their semi-major axes \citep[e.g.][]{2015ApJ...806...88S}. Eccentricity pumping in retrograde binaries may be fairly generic, since drag-like forces, such as dynamical friction, affect the binary angular momentum more strongly than the binary energy when applied near apocenter due to the lower velocities and larger lever arm. 

We have found that retrograde binaries contract $\approx 3-4$ times faster than prograde binaries for a given 
initial binary eccentricity. Similar to previous results (e.g., \citetalias{Dempsey2022}; \citealt{RLi2022}; \citetalias{dittmann2023}), we 
have also found that the orbital eccentricity of prograde binaries will decay. Since BBHs formed in the dynamical 
environment of the AGN disk are expected to form with high orbital eccentricities \citep[e.g.][]{JLi2023, whitehead2023, rowan2023}, 
understanding the rate of orbital decay versus eccentricity damping or pumping is important to make inferences 
about the binary when it enters GW detector bands. Prograde binaries have their eccentricities damped very slowly 
compared with their contraction rate when $e_b \lesssim 0.3$, where $\dot{a_b}/a_b \gtrsim 40 \dot{e_b}^2$.
Notably, the general eccentricity damping observed in this work differs from the current picture of isolated eccentric binary evolution, which suggests that binaries are driven towards some equilibrium nonzero eccentricity \citep[e.g.][]{2021ApJ...909L..13Z,2021ApJ...914L..21D,2023MNRAS.522.2707S}.

Retrograde binaries, on the other hand, grow their eccentricity much faster than prograde binaries damp their eccentricity. 
However relatively speaking, the eccentricity growth is quite slow compared with the orbital decay when $e_b \lesssim 0.5$, 
and becomes comparable in magnitude when $e_b \approx 0.7$. Given the binaries form with very high eccentricities \citep[$e_b \gtrsim 0.7$][]{JLi2023, rowan2023}, 
retrograde binaries should remain highly eccentric as their orbits decay, while prograde binaries should experience eccentricity 
damping which becomes inefficient when $e_b \approx 0.3$. Recently, \cite{RLi2023b} studied prograde BBHs in 2D with differing binary separation. Their Figure 7 shows that the torques inside the CSDs of the BBHs does not depended strongly on the binary separation, at least down to separations $\sim 3\times$ smaller than ours. However we note that they remove contributions from inside the sink region, which we do not do in our work. Since most of our gravitational torques manifest within the CSDs, it is plausible that our trends will remain the same for smaller binary separations. Assuming these trends remain similar as the binary separation decreases, 
we then expect that retrograde BBH mergers will have much larger eccentricities than prograde mergers.

Our result that retrograde binaries grow their eccentricity may alleviate several problems with the BBH in AGN disk GW channel. Firstly, previous works have shown that prograde binaries contract slowly, requiring several dozen mass doubling times to even come close to the GW emission regime. Furthermore, their contraction also appears to stall when an isothermal fluid is assumed (\citealt{YLi2021}; \citetalias{Dempsey2022}), although this is mitigated somewhat by assuming a non-uniform temperature profile \citep{YLi2022, RLi2022, RLi2023}. Eccentricity growth can solve this problem since the binary semi-major axis does not need to contract down several order of magnitude. Rather, large eccentricities can produce arbitrarily close pericenter passages, bringing the binary members close enough together for GWs to drive them quickly to coalescence within a faction of a mass-doubling timescale. Since their eccentricity is damped, prograde binary mergers cannot benefit from this mechanism. Given that most bound BBHs formed in the AGN disk are likely to be retrograde \citep[e.g.][]{JLi2023}, we then expect that most GW events from the AGN disk channel will come from high eccentricity retrograde progenitors, with a smaller contribution from prograde low eccentricity binaries.

Although our results suggest AGN BBH mergers should mostly originate from highly eccentric retrograde BBHs, their orbital eccentricity will be damped by GW emission \citep{peters1964}. How large the eccentricity will be at merger is not yet known. However GW emission reduces binary eccentricity during pericenter passages, while we have shown that most of the eccentricity pumping from the gas occurs at apocenter. How these two effects balance each other will determine the final BBH eccentricity at merger.

\subsection{Precession Evolution}
As shown in Figure \ref{fig:orbital_evo} and Table \ref{tab:com} binary-disk interactions induce an appreciable rate of precession in most binaries, on roughly the order of a radian per mass-doubling timescale. This disk-induced precession has the potential to disrupt certain dynamical processes which might otherwise be relevant to the orbital evolution of binary black holes orbiting SMBHs. Namely, the evection resonance is able to drive binaries to high eccentricities if $\Omega_0$ and $\dot{\varpi}$ are commensurable \citep[e.g.,][]{1994AJ....108.1943T,bhaskar2022,munoz2022}, and von Zeipel-Lidov-Kozai (ZLK) cycles may spur inclined binaries to extremely high eccentricities on a timescale of $\sim \Omega_b/\Omega_0^2$ \citep[][see also, e.g., \citealt{2019MEEP....7....1I} for a review]{1910AN....183..345V,1962P&SS....9..719L,1962AJ.....67..591K}. 

It is well-established that general relativistic precession precludes ZLK oscillations if relativistic precession occurs on a timescale much shorter than the ZLK timescale \citep{1997Natur.386..254H}, and that in more mild cases additional precession limits the maximum eccentricities attainable during ZLK cycles \citep[e.g.][]{2002ApJ...576..894M}. In the following we consider the importance of disk-induced precession relative to that driven by tidal interactions with the central SMBH to gauge the potential effect of the accretion disk on such processes. 

For simplicity, we parameterize the precession rate due to tidal interactions with the SMBH as 
\begin{equation}
\dot{\varpi}_{\bullet} = \mathcal{B}\Omega_0. 
\end{equation}
Similarly, we parameterize the effect of the AGN disk on binary precession according to 
\begin{equation}
\dot{\varpi}_{\rm gas} = \mathcal{A}\frac{\dot{m}_b}{m_b}, 
\end{equation}
where values for $\mathcal{A}$ and $\mathcal{B}$ may be read off of Table \ref{tab:com}. In the $a_b\ll R_H$ limit, one may approximate $\mathcal{B}$ analytically \citep[e.g.][]{2015MNRAS.447..747L,RLi2022}, which is in practice a fairly good approximation for the prograde binaries studied here. 

For the sake of comparison, we can parameterize the accretion rate as some multiple of the Eddington-limited accretion rate, such that $\dot{m}_b = \eta\dot{M}_{\rm Edd},$ where

\begin{equation}
\dot{M}_{\rm Edd} = \frac{4\pi G m_b}{\kappa c \epsilon} \approx 2.2\times10^{-6}\left(\frac{0.1}{\epsilon}\right) \left(\frac{m_b}{100\,M_\odot}\right)\,\rm{M_\odot\,yr^{-1}},
\end{equation}
where $\epsilon$ is the fraction of rest mass radiated during accretion onto the black holes, $\kappa$ is the opacity of the ambient gas, and $c$ is the speed of light.

In this case the relative importance of tidally-driven and disk-driven precession can be quantified through 
\begin{equation}
\frac{\dot{\varpi}_{\rm gas}}{\dot{\varpi}_\bullet}=\frac{\mathcal{A}}{\mathcal{B}\Omega_0}\frac{4\pi\eta G}{\kappa\epsilon c}.
\end{equation}
Assuming a fiducial hydrogen electron scattering opacity $\kappa=0.4\,{\rm cm^2\,g^{-1}}$, this is approximately 
\be
\frac{\dot{\varpi}_{\rm gas}}{\dot{\varpi}_\bullet} &\approx& \frac{\mathcal{A}}{\mathcal{B}}\!\left(\frac{\eta}{10^5}\right)\!\!\left(\frac{0.1}{\epsilon}\right)\!\!\sqrt{\!\!\left(\!\frac{10^6\,M_\odot}{M_\bullet}\!\right)\!\!\!\left(\frac{R_0}{0.1\,{\rm pc}}\right)^{3}}\!\!\!.
\ee

Notably, we measure accretion rates onto the binary that are a few percent of the Bondi accretion rate. 
Following \citetalias{dittmann2023}, the Bondi accretion rate relative to the Eddington accretion rate is approximately
\begin{equation}
\begin{split}
\frac{\dot{M}_B}{\dot{M}_{\rm Edd}} \approx 5\times10^5 
\left(\frac{R_H}{H}\right)^3\left(\frac{\rho_0}{10^{-14}\,\rm{g\,cm^{-3}}}\right)
\\\left(\frac{M_\bullet}{10^6\,M_\odot}\right)^{-1/2}
\left(\frac{R_0}{0.1 {\rm pc}}\right)^{3/2}\left(\frac{\epsilon}{0.1}\right),
\end{split}
\end{equation}
motivating our fiducial value of $\eta$. Thus, depending on the properties of the accretion disk around the binary, disk-induced precession varies from a small perturbation to a dominant driver of binary evolution. Notably, the fraction-of-Bondi accretion rates measured in our simulations could in principle be highly super-Eddington, so feedback processes may strongly affect the dynamics of these systems. 

\subsection{Circumsingle Disks}

\begin{figure}
    \includegraphics[width=1.0\linewidth]{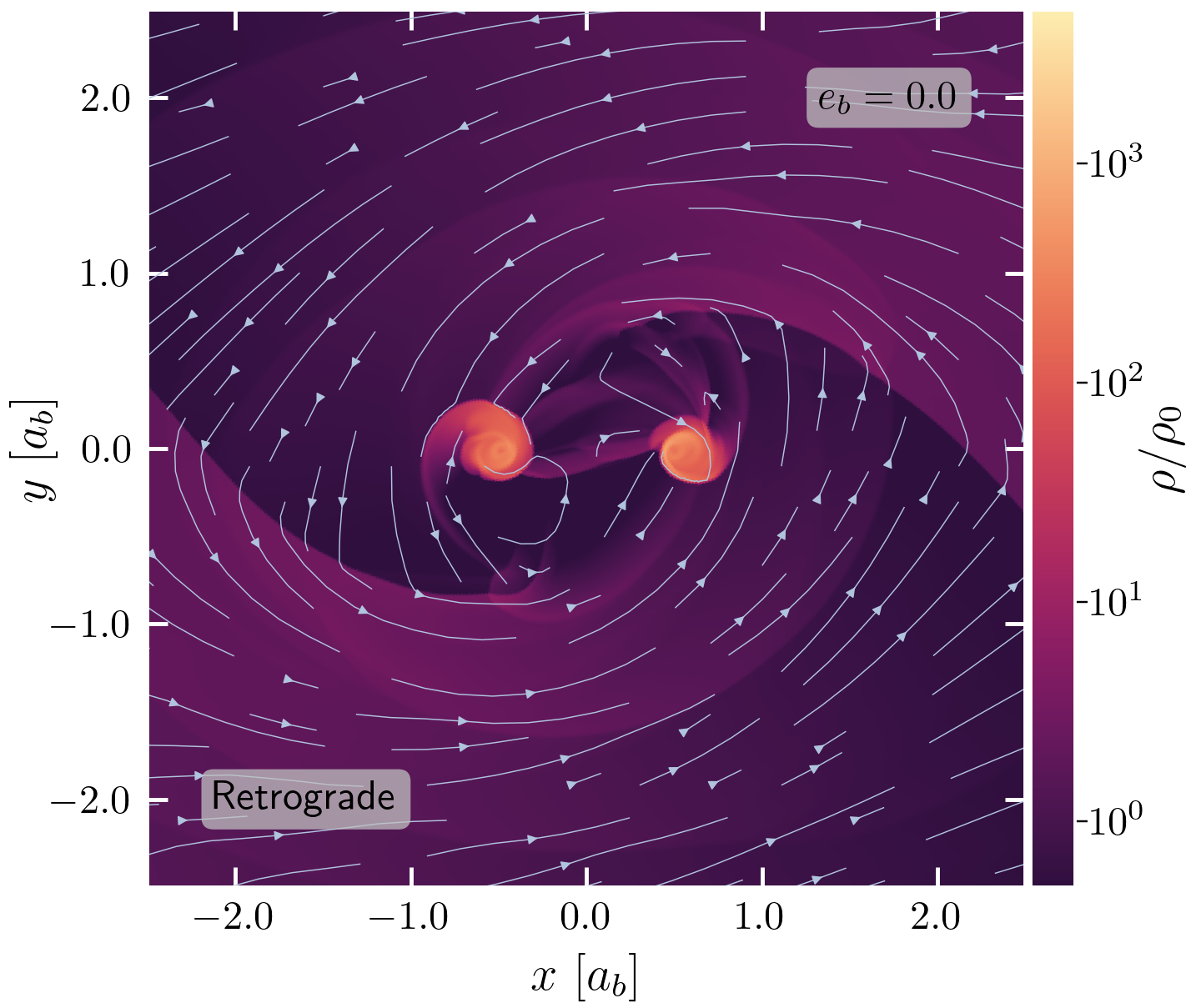}
    \caption{Surface density of our 2D retrograde $e_b=0$ simulation. CSDs form around the retrograde BHs and orbit in the same direction as the BBHs. }
    \label{fig:csd_slices2d}
\end{figure}

Our results indicate that gravitational interaction with the CSDs impart most of the torque and power onto the binary 
This is in agreement with the 3D circular and prograde results from \citetalias{Dempsey2022}. 
Given the importance of the CSD region, the sink prescription will have a significant effect on the balance of torque and power. In this work, as in our past work, we use torque-free sinks since the true accretion radius of the BHs is several orders of magnitude smaller than our sink region \citep{dempsey2020, dittmann2021}. A naive sink treatment would accrete angular momentum onto the BHs which would be far larger than the true value at the BH innermost stable circular orbit. This will result in a large excess of angular momentum being taken from the CSD region, while its structure is altered \citep{dittmann2021}. This is likely occurring in the BBH formation and merger simulations presented in \cite{rowan2023}, which used the SPH code \texttt{phantom} with the default sink prescription \citep{phantom2018}.
Given these results, a resolution study with smaller sink radii is necessary.
However, the computational cost of such a study would be extremely high if one wished to maintain, e.g., a constant number of cells per sink radius, and so we defer answering this question to a future study.

We observe CSDs in our retrograde simulations, although they were not seen in the 2D retrograde and circular binary simulation by \cite{RLi2022}. By running a 2D retrograde simulation, we verified that CSDs still form (Figure \ref{fig:csd_slices2d}). 
The morphology of the CSDs we see in 2D retrograde simulations is similar to the CSDs in 3D. The difference between our 2D simulation and the one presented in \cite{RLi2022} that is generating this discrepancy is likely to be the equation of state. In our simulations the gas suddenly transitions from prograde around the binary to retrograde in the CSD. For a non-isothermal equation of state this will lead to a sudden increase in the temperature of the gas, which might partially or totally inhibit the creation of the CSDs. The CSDs around the retrograde BBHs are also retrograde, as they also are in 3D. \cite{tiede2023} simulated isolated BBHs that are retrograde with respect to their surrounding CBD in 2D. They found that the CSDs around the individual BHs orbit in the same direction as the CBD, which is opposite to our finding for BBHs in an AGN disk. Given our finding that the CSDs orbit in the same direction as the BBHs in 2D, this difference probably arises due to the lack of a clear CBD and heavily depleted cavity in our simulations.

\subsection{Limitations}
There are several important limitations and caveats to our results. Firstly, we have assumed an inviscid, isothermal fluid devoid of magnetic fields. It has been shown by other works that changing the equation of state away from isothermal generally produces faster mergers across a larger range of binary parameters \citep{YLi2022, RLi2023b}. Similarly, the super-Eddington accretion onto each BH may produce substantial feedback, causing actual accretion flows to strongly deviate from those studied here. Secondly, although we have produced some of the highest-resolution 3D simulations of BBHs in AGN disks to date, our results may still be resolution and sink dependent. Our results show that the most important torque and power contribution originate within the CSDs and close to the sinks. Simulations with a more sophisticated treatment of the thermodynamics, at higher resolution, and with smaller sinks, should be performed to test whether our results, particularly with regards to eccentricity pumping, still hold.

\section{Conclusions}\label{sec:sum}
We performed a series of 3D hydrodynamical simulations of prograde and retrograde eccentric BBHs embedded in an AGN disk. We have shown that retrograde eccentric BBHs \emph{increase} their orbital eccentricity over time while decreasing their separation. The increase is monotonic with the binary eccentricity. This means that retrograde BBHs may experience a runaway growth of their orbital eccentricity leading to arbitrarily close pericenter passages of the BHs. These close passages can allow the BBHs to dissipate their orbital energy (and hence decrease their semi-major axis) through GW emission, rather than through the much more inefficient gas energy dissipation. This parallels the mechanism proposed for retrograde SMBH binary evolution proposed in \citet{2015ApJ...806...88S}. We have also found that retrograde binaries contract $3-4\times$ faster than their than prograde counterparts. Prograde binaries experience eccentricity damping, however this becomes inefficient when the binary eccentricity is $e_b \lesssim 0.3$. 

Our results, combined with \citet{JLi2023} and \citetalias{dittmann2023}, trace a pathway for high-mass black hole binary mergers
in AGN disks as a viable progenitor for GW events. 
Once a large number BHs are captured or formed within an AGN disk, they quickly have close encounters with other BHs \citep{2020MNRAS.498.4088M,tagawa2020}.
The dissipation of the close encounter energy by the gaseous medium facilitates the capture of the two BHs into a BBH \citep{JLi2023}. 
The properties of the resultant binary tend to be retrograde and eccentric with a separation that is a fraction of the mutual Hill radius \citep{JLi2023}. 
Once formed in such a configuration, we have shown that the likely outcome is the further increase of the BBH eccentricity, lowering the pericenter distance to a point where GW radiation takes over and quickly (relative to a circular binary) merges the BHs to produce a LIGO/VIRGO GW event. Furthermore, depending on the binary location within the AGN disk, disk-driven precession may dominate over tidally-driven precession, which may negate additional eccentricity pumping mechansisms such as evection resonances and ZLK cycles. 

\section*{Software}
\texttt{matplotlib} \citep{matplotlib}, \texttt{scipy} \citep{scipy}, \texttt{numpy} \citep{numpy}, \texttt{yt} \citep{yt}, \texttt{Athena++} \citep{stone2020}, \texttt{REBOUND} \citep{rein2012, rein2015}

\section*{Acknowledgments}
We gratefully acknowledge the support by
LANL/LDRD under project numbers 20200772PRD4 and 20220087DR.
This research used resources provided by the Los Alamos
National Laboratory Institutional Computing Program,
which is supported by the U.S. Department of Energy
National Nuclear Security Administration under Contract No. 89233218CNA000001. 
Approved for release under LA-UR-23-32647.

\appendix

\section{CSD Separatrix Fitting}

Our procedure for fitting the CSD separatrix is as follows. Firstly, we take $v_x$, $v_y$, and density values along the $z=0$ slice from our simulation snapshots. We then rotate and rescale these quantities so that they are now in the $\xi$ and $\eta$ pulsating and rotating coordinate system given by equations \ref{eq:eta} and \ref{eq:xi}. We then convert to cylindrical coordinates centered on each BH, and combine and average the resulting $v_r$, $v_\phi$, and density from the frame of each BH. This gives us an average of the velocity and density profile of the two CSDs in a single snapshot. We then average the snapshots around $\pm \pi/8$ of pericenter and apocenter. 

\begin{figure*}
    \centering
    \includegraphics[width=0.8\linewidth]{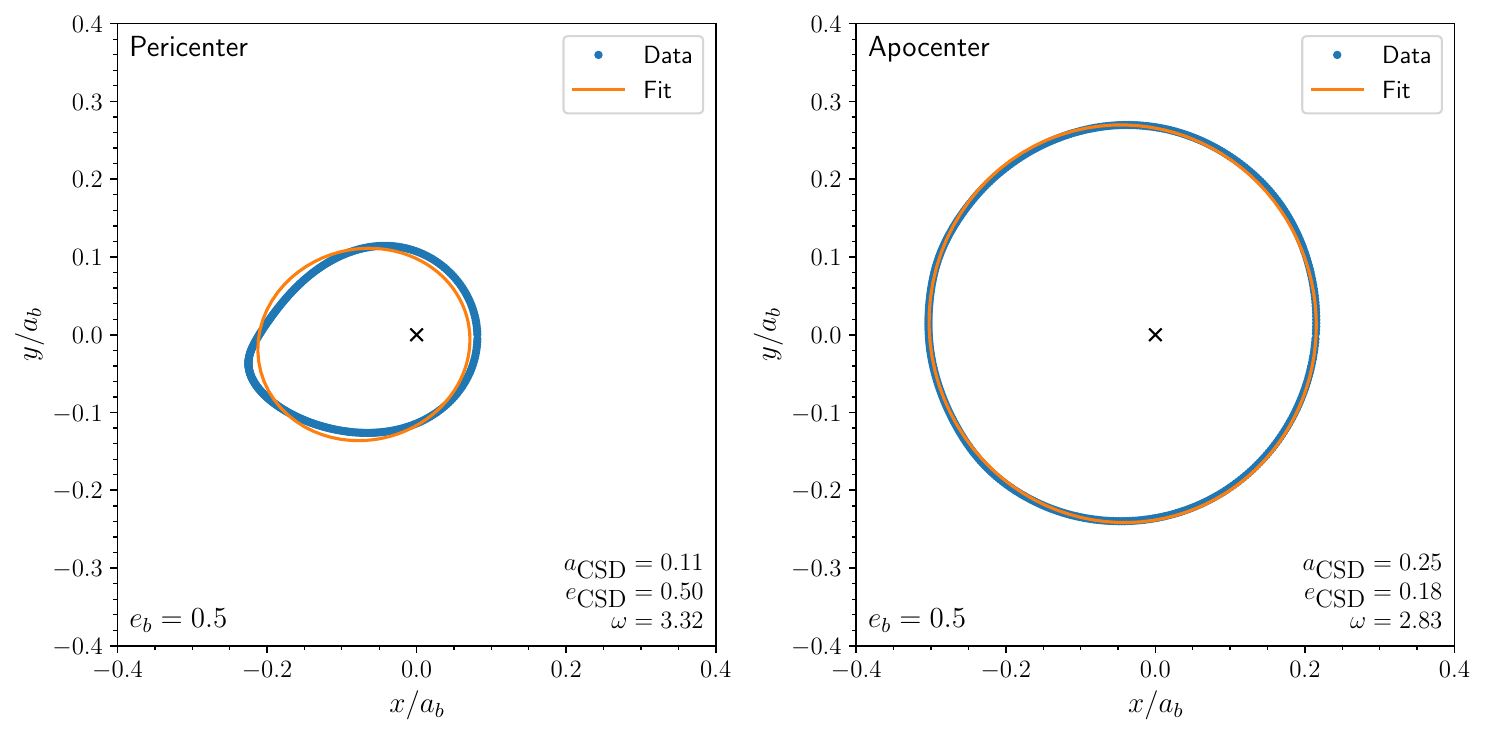}
    \caption{An example of our separatrix fitting procedure on our time-averaged $e_b = 0.5$ prograde simulation.}
    \label{fig:csd_ecc}
\end{figure*}

We find the separatrix in the time-averaged snapshots of our simulations by integrating velocity streamlines in the plane of the binary orbit ($z=0$) using a second-order Runge-Kutta method. We also constrain the method so that $dv_\phi/dr$ is positive (negative) for the prograde (retrograde) CSDs. We then fit the separatrix with an ellipse. Figure \ref{fig:csd_ecc} shows the CSDs in our prograde $e_b = 0.5$ simulation and the resulting fit of the ellipse.

We note that our definition of the CSD from the separatrix will not be applicable to 2D simulations. In 3D, the midplane is characterized by \emph{decretion}, while accretion happens on the surface layers. In 2D the flow is restricted to a thin sheet to which accretion is restricted and hence $v_r$ is negative everywhere in 2D accretion disks.

\bibliography{paper}{}
\bibliographystyle{aasjournalnolink}

\end{document}